\documentclass[conference]{IEEEtran}
\IEEEoverridecommandlockouts
\usepackage{cite}
\usepackage{amsmath,amssymb,amsfonts}
\usepackage{graphicx}
\usepackage{textcomp}
\usepackage{xcolor}
\usepackage{algorithm}
\usepackage{algorithmicx}
\usepackage{algpseudocode} 
\usepackage{diagbox}
\usepackage{multirow}
\usepackage{subfigure}
\usepackage{enumitem}
\usepackage{url}
\usepackage{marvosym}
\def\BibTeX{{\rm B\kern-.05em{\sc i\kern-.025em b}\kern-.08em
    T\kern-.1667em\lower.7ex\hbox{E}\kern-.125emX}}
\begin{document}
\title{FOSS: A Self-Learned Doctor for Query Optimizer}
% {\footnotesize \textsuperscript{*}Note: Sub-titles are not captured in Xplore and
% should not be used}
% \thanks{Identify applicable funding agency here. If none, delete this.}
% \author{\IEEEauthorblockN{Kai Zhong, Luming Sun,Tao Ji, Cuiping Li, Hong Chen}
% \IEEEauthorblockA{\textit{Key Laboratory of Data Engineering and Knowledge Engineering, Ministry of Education} \\
% \textit{\textit{School of Information, Renmin University of China}}\\
% % Beijing, China \\
% zhongkai1203, jitao, licuiping, chong@ruc.edu.cn}
% }
\author{
\IEEEauthorblockN{
    Kai Zhong$^{\dagger *}$, 
    Luming Sun$^{\mathsection \dagger}$, 
    Tao Ji$^{\dagger *}$,  
    Cuiping Li\textsuperscript{\Letter}$^{{\dagger *}}$,
    Hong Chen$^{{\dagger \ddagger}}$\\}
\IEEEauthorblockA{\textit{Renmin University of China, China$^{\dagger }$} \\
\textit{Key Laboratory of Data Engineering and Knowledge Engineering, MOE, China$^{*}$}\\
\textit{Engineering Research Center of Database and Business Intelligence, MOE, China$^{\ddagger}$}\\
\textit{Shanghai Yunxi Technology Co., Ltd.$^{\mathsection}$}\\
\hspace{0.3em} 
\{zhongkai1203,\hspace{0.2em}jitao,\hspace{0.2em}licuiping,\hspace{0.2em}chong\}@ruc.edu.cn,\hspace{0.2em}sunluming@inspur.com}
 \IEEEcompsocitemizethanks{
  \IEEEcompsocthanksitem \Letter~Corresponding author
  \IEEEcompsocthanksitem This is the accepted version of the paper published in ICDE2024, DOI: 10.1109/ICDE60146.2024.00330. The final published version is available at https://ieeexplore.ieee.org/abstract/document/10597900.
 }
}
\maketitle
% \author{
% {\IEEEauthorblockN{
%     Kai Zhong\IEEEauthorrefmark{1}\IEEEauthorrefmark{3},
%     Luming Sun\thanks{ Cuiping Li is the corresponding author.}\IEEEauthorrefmark{2}, 
%     Tao Ji\IEEEauthorrefmark{1}\IEEEauthorrefmark{3},
%     Cuiping Li\IEEEauthorrefmark{1}\IEEEauthorrefmark{3},
%     Hong Chen\IEEEauthorrefmark{1}\IEEEauthorrefmark{3}}
%     \IEEEauthorblockA{\IEEEauthorrefmark{1}xxx, China, \href{mailto:xxx@xxx,xxx@xxx}{\{xxx, xxx\}@xxx}\\
% \IEEEauthorrefmark{3}xxx, China, \href{mailto:xxx@xxx,xxx@xxx}{\{xxx, xxx\}@xxx}\\
% \IEEEauthorrefmark{2}xxx, China,  \href{mailto:xxx@xxx}{xxx@xxx}}
% }}
    % \IEEEauthorblockN{$^1$Kai Zhong, $^2$Luming Sun, $^1$Tao Ji, $^{1,*}$Cuiping Li, $^1$Hong Chen}
    % \IEEEauthorblockA{\textit{School of Information, Renmin University of China, Beijing, China}}
    % \IEEEauthorblockA{\textit{Key Laboratory of Data Engineering and Knowledge Engineering, MOE, China}}
    % \IEEEauthorblockA{\textit{$^2$ , , Beijing, China}}
    % \IEEEauthorblockA{zhongkai1203, sunluming, jitao, licuiping, chong@ruc.edu.cn}
% \author{\IEEEauthorblockN{Tao Ji, Luming Sun, Cuiping Li, Hong Chen}
% \IEEEauthorblockA{\textit{Key Laboratory of Data Engineering and Knowledge Engineering, Ministry of Education} \\
% \textit{\textit{Renmin University of China}}\\
% % Beijing, China \\
% }
\begin{abstract}
Various works have utilized deep learning to address the query optimization problem in database system. They either learn to construct plans from scratch in a bottom-up manner or steer the plan generation behavior of traditional optimizer using hints. While these methods have achieved some success, they face challenges in either low training efficiency or limited plan search space. To address these challenges, we introduce FOSS, a novel framework for query optimization based on deep reinforcement learning. FOSS initiates optimization from the original plan generated by a traditional optimizer and incrementally refines suboptimal nodes of the plan through a sequence of actions. Additionally, we devise an asymmetric advantage model to evaluate the advantage between two plans. We integrate it with a traditional optimizer to form a simulated environment. Leveraging this simulated environment, FOSS can bootstrap itself to rapidly generate a large amount of high-quality simulated experiences. FOSS then learns from these experiences to improve its optimization capability. We evaluate the performance of FOSS on Join Order Benchmark, TPC-DS, and Stack Overflow. The experimental results demonstrate that FOSS outperforms the state-of-the-art methods in terms of latency performance. Compared to PostgreSQL, FOSS achieves speedup ranging from 1.15x to 8.33x in total latency across different benchmarks.
\end{abstract}


\section{Introduction}
\label{Intro}
Query optimizer plays a critical role in database management system (DBMS). It takes a relational algebra expression derived from the parser as input and outputs an optimized physical execution plan. Most query optimizers follow the Selinger's design and consist of the cardinality estimator, the cost estimator, and the query plan generator \cite{3part}. These components work together to explore the plan space and find the potentially optimal query plan.

However, for traditional query optimizer, finding the optimal plan within the vast plan space is an NP-hard problem \cite{JOB}. Even if only the left-deep tree is considered in join order, the plan space still grows exponentially to $O(n!)$, where $n$ represents the number of tables involved in the query. The plan search space is further expanded when considering other factors, such as bushy tree, join methods and access path. Meanwhile, errors in traditional cardinality and cost estimator often result in poor-performance plans \cite{JOB, MOSE}. Furthermore, traditional query optimizer can not learn from the past experiences, often generating the similar poor-performance plans for the same queries repeatedly \cite{DBAIsurvey}.

\subsection{Machine Learning for Query Optimization}
To address the problem of traditional query optimizer, numerous studies focus on employing deep learning in end-to-end query optimization. We roughly categorize current methods into two types: plan-constructor and plan-steerer.\\
\textit{\underline{Plan-constructor.}} It is an \textbf{expertise-omission} approach that discards or underuses expert knowledge of traditional optimizer and focuses on constructing plans from scratch \cite{Rejoin,DQ,Tree-lstm,Neo,Balsa,LOGGER}. It models the process of query plan generation as a Markov Decision Process (MDP) and uses deep reinforcement learning (DRL) to solve it. Its framework involves several components. An agent (i.e., a learned model) starts with a set of individual tables involved in the query, and at each step, it joins tables, joins subplans or specifies an operator. Each step results in a new state, which consists of the partial plan and other tables that have not been joined yet. The process continues until all tables involved in the query are joined, resulting in a complete plan. In this process, the DBMS serves as the environment and provides reward feedback, which are typically correlated with the execution cost or true execution latency of the final plans. Although plan-constructor can gradually surpass the performance of traditional query optimizer after a certain training period, it has the following drawbacks:
\begin{itemize}%[leftmargin=0.3cm, itemindent=0.5cm]
\item \textbf{(C1) Learning from scratch.} The agent need to exert considerable effort to learn how to produce a cost-effective plan from scratch. This usually requires a significant amount of trial-and-error learning, gradually enabling the agent to reach a proficiency level similar to that of a traditional query optimizer.
% \item \textbf{Long sequence decision.} The number of steps required for an agent to complete the construction of a plan depends on the number of tables involved in the query. In Join Order Benchmark (JOB) \cite{JOB}, the queries have between 3 and 16 joins, with an average of 8 joins per query. To optimize decisions with such a long sequence of steps is a challenging problem for DRL.
\item \textbf{(C2) Poor experience sampling.} Training a proficient agent requires a substantial amount of high-quality experiences. Prior approaches \cite{Rejoin,DQ,Tree-lstm,Neo,LOGGER,Balsa} commonly employ cost estimation or execution latency as rewards to shape the experiences. Nevertheless, errors in cost estimation arising from the traditional cost model frequently lead to suboptimal outcomes. While using execution latency is more likely to generate superior plans, the low efficiency of obtaining execution latency incurs significant overhead. This inefficiency leads to the insufficient exploration problem, thus requiring a large amount of time for model convergence.
% \item \textbf{Sparse reward.} Prior methods such as \cite{Balsa,Neo,Tree-lstm,JOGGER} encounter challenges in providing specific evaluations for subplans before reaching the final state. In such scenarios, rewards are often assigned a value of 0 for all steps except the final one, leading to sparse rewards for the agent and limiting access to sufficient guidance \cite{sparse_rewards}. %The challenge of sparse rewards hampers the agent's learning, limiting its access to sufficient guidance \cite{sparse_rewards}.\\
 \item \textbf{(C3) Sparse reward.} The number of steps required for an agent to complete the construction of a plan depends on the number of tables involved in the query. In Join Order Benchmark (JOB) \cite{JOB}, the queries have between 3 and 16 joins, with an average of 8 joins per query. However, prior methods such as \cite{Neo,Tree-lstm,JOGGER} encounter challenges in providing specific evaluations for subplans before obtaining the complete plan. In their scenarios, rewards are often assigned a value of 0 for the intermediate state, leading to sparse rewards for the agent and limiting access to sufficient guidance \cite{sparse_rewards}.
\end{itemize}
\textit{\underline{Plan-steerer.}} It is a \textbf{blackbox-expertise} approach that leverages the embedded expert knowledge in traditional query optimizer \cite{Bao,Lero,HybridQO, DeepO}. Guided by various hints, it generates multiple plans through the traditional query optimizer. For instance, in Bao \cite{Bao}, coarse-grained hints like disabling nested loop join for the entire query are used to guide the traditional query optimizer. It then utilizes a learned value network to evaluate these candidate plans and predict the cost. Ultimately, it selects the plan with the potentially lowest cost for execution. However, it has the following problems:
\begin{itemize}%[leftmargin=0.3cm, itemindent=0.5cm]
\item \textbf{(S1) Requiring expertise for hint design.} While plan-steerer leverages traditional query optimizer to generate better plans through hints guidance, the design of hints still requires expert knowledge, as seen in the case of the hint set grouping in Bao \cite{Bao}.
\item \textbf{(S2) Limited search space.} While some hints can steer traditional optimizer to select better plans, the coarse-grained nature of hints may limit plan-steerer to settling for the suboptimal plans.
\item \textbf{(S3) Hard to balance.} Adding more candidate hints increases the number of candidate plans, thereby raising the probability of generating a better plan. However, this expansion also leads to an increase in optimization time.
% Determining the appropriate size of candidate hints is a question that requires further investigation.
\end{itemize}

\subsection{A Novel Framework for Query Optimization}
In this paper, to alleviate the aforementioned problems, we introduce a novel framework \textbf{FOSS} that modi\underline{\textbf{f}}y \underline{\textbf{o}}riginal plan \underline{\textbf{s}}tep by \textbf{\underline{s}}tep. FOSS can be categorized as \underline{\textit{plan-doctor}}, which is a \textbf{whitebox-expertise} approach. Unlike the blackbox-expertise approach that steers or constrains the traditional optimizer behavior, FOSS leverages expert optimization knowledge more explicitly by optimizing the plans generated by traditional query optimizer. 

The insight behind FOSS is that although traditional query optimizer often produces suboptimal plans due to errors in cardinality estimation and cost estimation, much better plans can be retrieved by making a limited number of modifications to these suboptimal plans. For example, Query 1b from JOB takes $100.67ms$ to execute with plan generated by PostgreSQL's optimizer. Due to the inaccurate cost estimation, the optimizer chooses a hash join operator between table $it$ and table $mi\_idx$, which slows down the execution of plan. In this example, FOSS will first override the join method between the two tables to use a nested loop join, then exchange the positions of $it$ and $mi\_idx$ to a proper order. With FOSS, the total execution latency reduces to $0.27 ms$. In similar situations, FOSS acts like a doctor, identifying the suboptimal aspects causing performance issues in the given plan. It then takes actions, such as changing the physical implementation of the join node or rendering a proper order for two tables, in a step-by-step manner to optimize it.
% However, simply changing the physical implementation of the join node and rendering a proper order of the two tables result in much lower execution latency. %To identify the above issues and make corrections, we build a framework called \textit{\textbf{plan-doctor}}.

% Version 11_26:At a high level, FOSS treats the aforementioned process as a MDP and also employs DRL to solve it. FOSS aims to learn an agent to optimize the original plan generated by traditional query optimizer with a limited number of steps and output the optimal plan from all candidate plans generated during optimization process. To achieve this, FOSS also learns an \textit{asymmetric advantage model} (AAM), which evaluates the advantage between a pair  of candidate plans. Following the temporal sequence (i.e., the order of generated candidate plans), FOSS assesses specific pairs of candidate plans and selects the estimated optimal plan with the help of AAM. Furthermore, FOSS integrates AAM with a traditional query optimizer to create a \textit{simulated environment}. This integration allows FOSS to efficiently bootstrap itself and enhance its optimization capability with high-quality simulated experiences. 
%Version 11_27
However, to obtain an excellent plan-doctor, two key capabilities are essential: \romannumeral1) the ability to identify suboptimal aspects in the original plan and generate appropriate candidate action sequences for optimization. Note that different action sequences applied to the original plan will result in distinct new plans. \romannumeral2) the ability to select the sequence that yields the optimal result for the original plan from candidates. To achieve these objectives, for \romannumeral1), FOSS learns a planner that models the identification of suboptimal aspects and the generation of action sequences as a MDP and employs DRL to solve it, and for \romannumeral2), FOSS learns an asymmetric advantage model (AAM) to serve as the optimal candidate action sequence selector. AAM evaluates advantage values between pairs of candidate plans generated by candidate action sequences to determine the optimal plan among them. Furthermore, inspired by model-based reinforcement learning (MBRL), FOSS integrates the AAM with a traditional optimizer to form a simulated environment, where the AAM serves as a reward indicator, and the traditional optimizer acts as a state transitioner.

We summarize the features of FOSS as follows:
\begin{itemize}
\item\textbf{High training efficiency.} FOSS starts from the original plans and focuses on optimizing suboptimal aspects, eliminating the need to learn from scratch (addressing \textbf{C1}). At each step, FOSS can generate a new complete execution plan, allowing for a straightforward evaluation of reward based on the performance of the new plan. Furthermore, the experimental results in \ref{experiments} demonstrate that FOSS often requires only 1-3 steps to achieve a better-performance plan (addressing \textbf{C3}). With the simulated environment, FOSS can efficiently bootstrap itself, generating ample high-quality simulated experiences to enhance its optimization capability (addressing \textbf{C2}).
% For instance, rewards can be evaluated at a low cost based on the advantage between plan of each step and previous optimial plan. (addressing C2 and C3).
\item \textbf{Intelligent candidate plans generation.} FOSS can autonomously generate candidate plans through learning a planner, unlike plan-steerer's reliance on pre-injected expert knowledge for hint design (addressing \textbf{S1} and \textbf{S3}).
\item \textbf{Sufficient plan search space.} FOSS employs finer-grained optimization, in contrast to plan-steerer's coarse hints. Under similar scenarios, FOSS shares the same plan search space as plan-constructor, indicating its potential to discover the globally optimal plan (addressing \textbf{S2}).

\end{itemize}
    
    Neither plan-constructor nor plan-steerer alone can encompass all of the aforementioned features. Therefore, FOSS strikes a good balance between the two approaches.
\begin{figure*}
  \centering
  \includegraphics[width=\linewidth ]{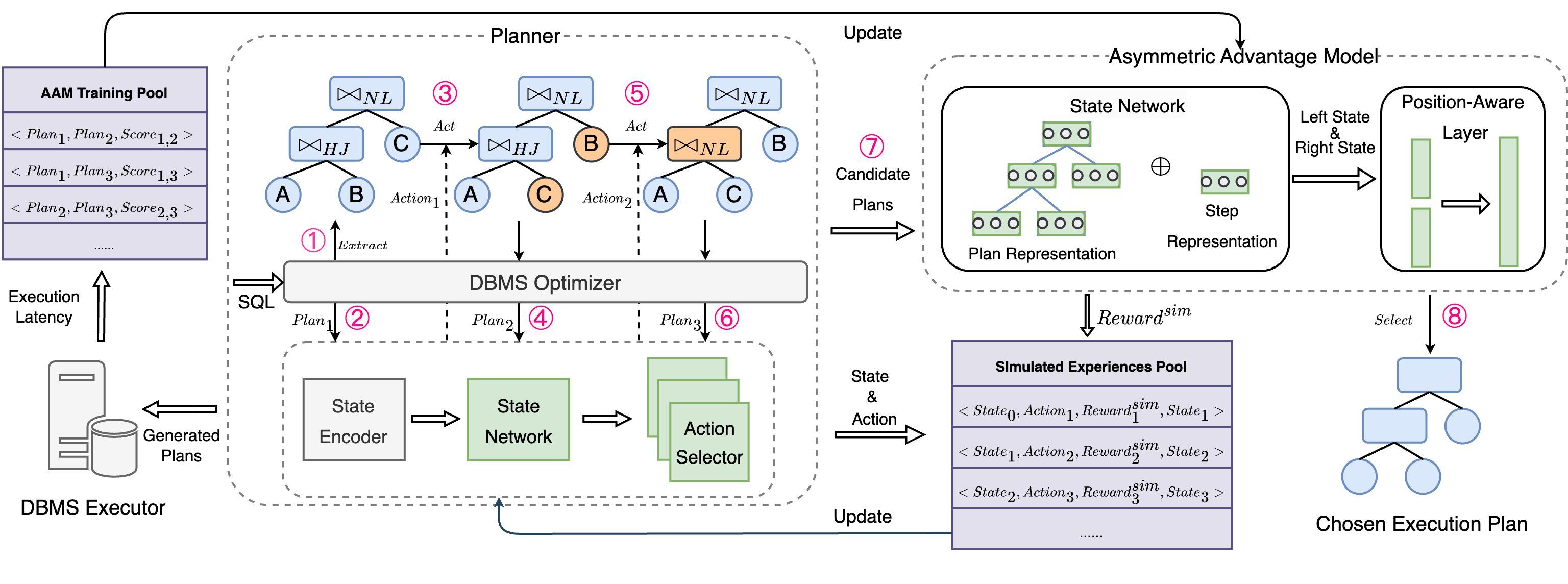}
  \caption{The framework of FOSS.}
  \label{fig:overview}
\end{figure*}

% To train an excellent plan-doctor, a significant amount of high-quality experience data is required. Every step of the plan-doctor may yield a new plan, and the assessment of this new plan is then used as feedback to provide reward to the agent, aiding its decision-making and updates. Using the execution latency or cost estimation of plan as feedback is a natural approach, but as mentioned earlier, there is a challenge in balancing effectiveness and efficiency. Inspired by model-based RL \cite{MBRL}, we utilize the asymmetric advantage model along with the DBMS optimizer to construct a \textbf{\textit{simulated environment}} that facilitates efficient interaction with plan-doctor. It can generate a greater amount of high-quality experiences, which significantly expedites the learning of plan-doctor.
\subsection{Contributions}
We summarize our contributions as follows:
\begin{itemize}
\item  We introduce a whitebox-expertise approach called plan-doctor, which consists of a planner and a selector. It produces better-performance plans by optimizing the original plan through a step-by-step process.
\item We construct a simulated environment and use simulated experiences to accelerate the learning of FOSS.
\item We design an asymmetric advantage model to serve as a reward indicator in the simulated environment and as a selector to choose the optimal plan among candidate plans.
\item Experimental results show that FOSS outperforms the state-of-the-art methods in terms of latency performance, while also achieves speedup ranging from $1.15\times$ to $8.33\times$ in total latency  compared to the PostgreSQL.
\end{itemize}

\section{SYSTEM OVERVIEW}
\begin{figure*}[h]
  \centering
  \includegraphics[width=\linewidth]{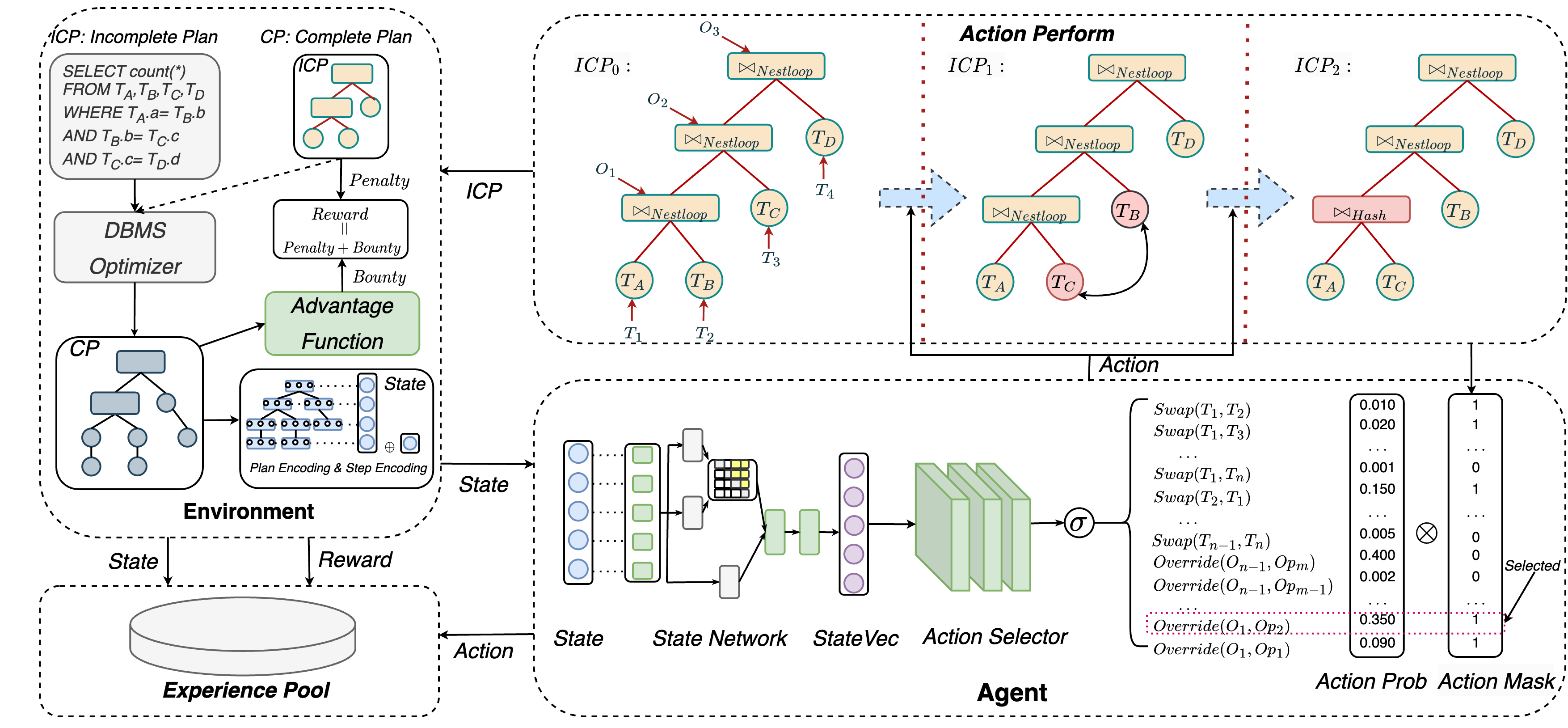}
  \caption{The workflow of planner.}
  \label{fig:plan-step}
\end{figure*}
The overall architecture of FOSS is illustrated in Fig. \ref{fig:overview}. For a given query, FOSS follows the order indicated by the pink numbers to output the better-performance plan. Firstly, \textbf{planner} utilizes past experiences to iteratively modify the original plan, resulting in a set of potentially promising plans. Then, following the temporal sequence (i.e., the order of generated candidate plans), \textbf{asymmetric advantage model} (AAM) serves as the selector, assessing specific pairs of candidate plans and selecting the estimated optimal plan. To expedite the learning of the planner, FOSS builds a \textbf{simulated learner} to generate ample high-quality experiences. We divide our framework into the following three key modules.\\
% To expedite the effective training of the plan-doctor, FOSS builds a simulated environment to generate ample high-quality experiences.
%In general, it observes the plan, conducts an analysis, and proposes a recommended action sequence for making optimizations.
%It is an abstract approach based on DRL for query optimization, and it can be regarded as the cornerstone of FOSS.
\textbf{Planner.} We deploy the entire process of planner using DRL framework. Starting from the original plan, planner first encodes plan features. Considering the temporal nature of plans generated by planner, we incorporate the step status. These features are concatenated as the initial state and then passed to the agent. The agent first represents the state using a state network. Then the state representation is subsequently forwarded to the action selector to generate an action, which could involve operations such as swapping table positions or overriding join method. Following the execution of the action, a new plan is generated and forwarded to the reward function module to obtain rewards. Meanwhile, the new plan is utilized to generate a new state. This iterative process continues until the maximum step limit is attained. Experiences in the format $\{State, Action, Reward, State_{'}\}$ will be collected to update the agent. Through extensive training, planner can generate better plans than the original plan. \\
% The plan-doctor operates within the same search space as plan-constructor, indicating that plan-doctor also has the potential to discover the globally optimal plan. However, on average, plan-doctor requires fewer steps to obtain a good enough plan compared to plan-constructor. This implies that plan-doctor is easier to train. Compared to plan-steerer, plan-doctor has a higher probability to generate high-quality candidate plans within a limited steps through training and it does not require expert knowledge to design specific hints.\\
\textbf{Asymmetric Advantage Model.} It serves as both optimal plan selector and reward indicator in simulated environment. It consists of state network and position-aware output layer. First, we define the state network, which also helps the planner's agent analyze the state. Its primary architecture is based on the Transformer \cite{Transformer}, which can efficiently capture the structural and node information of plans. It involves embedding plan features and step statuses separately, passing them through a multi-head attention network, ultimately outputting the state representations. For position-aware output layer, the input comprises two state representations $statevec_{left}$ and $statevec_{right}$, while the output indicates the extent to which $statevec_{right}$ surpasses $statevec_{left}$. We map the output to different scores. The selection of the optimal plan and the evaluation of rewards is determined based on these scores.\\
\textbf{Simulated Learner.} In DBMS, the cost of using the traditional optimizer to generate plans is low. FOSS utilizes the traditional optimizer and AAM to form a simulated environment. Therefore, FOSS comprises three key components: the agent, the real environment, and the simulated environment. During the training of FOSS, the agent will concurrently interact with both the real environment and the simulated environment. During the interaction with the real environment, the generated plans are executed in DBMS to obtain execution latency, forming the AAM training data pool. The data are then used to update the AAM through a supervised learning process. We also incorporate a dynamics timeout mechanism to ensure efficient execution latency collection. During the interaction with the simulated environment, a large amount of high-quality simulated experiences will be generated to update the agent. Meanwhile, we integrate a validation mechanism during the collection of simulated experiences. FOSS will collect the plans with high estimated performance, execute these plans in real environment to obtain their execution latency, and then incorporate them into the AAM training pool.
\section{Planner}
In this section, we will provide a detailed description of planner. Note that in current framework, we only consider left-deep plans, which is a structure frequently used in existing DBMS such as PostgreSQL and MySQL. Although we can explore bushy plans and incorporate modifications to the structure of plans in the action space, it significantly increases the size of the action space without necessarily yielding corresponding benefits. Therefore, for now, we focus solely on left-deep plans, leaving the consideration of bushy plans for future work.

Given a schema with $n$ tables, when a query $q$ is received, we obtain the complete plan $CP$ (i.e., execution plan) from the output of the traditional optimizer. As shown in the top right corner of Fig. \ref{fig:plan-step}, we extract the join methods and join order from $CP$, which significantly affect the performance of $CP$ \cite{LOGGER}. We refer to such a tree structure containing only the join order and join methods as the incomplete plan $ICP$. The planner's main process is to iteratively modify $ICP$ in a step-by-step manner. After taking one step on the $ICP$, guided by the new $ICP$, we can derive the new $CP$ using the DBMS extension, such as \textit{pg\_hint\_plan}\cite{pghintplan} in PostgreSQL. With the extension, table scan operators and other nodes will be complemented by the traditional optimizer using its own expert knowledge, resulting in the execution plan $CP$. When the maximum number of steps $maxsteps$ is reached, the plan modification process will be stopped. We refer to this process as an episode for an input query. Planner adopts the framework of DRL, which consists of key components like agent, environment, state, action and reward. Next, we will provide a detailed explanation of each component.\\
\textbf{State.}  In DRL, it is the output of the environment and the input to the agent. In planner, it is primarily composed of $CP$. To facilitate the subsequent processing by the agent, we pick up the key features from $CP$ and encode them into vectors. We will provide the specific details of $CP$ encoding in \ref{state repre}. Now, we use $PlanEncoding(CP)$ to represent the encoding of $CP$. Additionally, to incorporate the temporal characteristics of the state, we also include the $Step(t) = t/maxsteps$ in state. By concatenating these two types of features, we obtain the state features $State(CP_t)$ at time step $t$.
\begin{equation}\label{state}
State(CP_{t}) =  PlanEncoding(CP_{t}) || Step(t)
\end{equation}
\textbf{Action.} First, we assign sequential labels to the nodes of the $ICP$ in a bottom-up fashion. We use two types of labels: $T_{k}$ to represent leaf nodes (i.e., tables) and $O_{k}$ to represent non-leaf nodes (i.e., join methods). As shown in Fig. \ref{fig:plan-step}, we start with the two leaf nodes at the deepest level, where the left node is labeled as $T_{1}$ and the right node as $T_{2}$. The leaf node at the level above them are labeled as $T_{3}$, and so on. Similarly, we label the deepest non-leaf node as $O_{1}$, its parent node as $O_{2}$, and so forth, following the bottom-up sequence.

Next, we will explain the representation of two types of action. The first type is swapping the positions of two tables $(T_{l},T_{r})$, represented as $Swap(T_{l}, T_{r})$. There are $I_{s} = (n \times (n-1))/2$ different actions for table positions swapping. The second type is overriding the join method of node $O_{i}$ with the $j$-th join method in $Op$, represented as $Override(O_{i}, Op_{j})$, where $Op$ is the set of all available join methods in DBMS. There are $I_{o}=\left | Op \right | \times (n-1)$ different actions for overriding join method. Action is encoded as an integer $a$ within the range $[1, I_{s}+I_{o}]$. We define an action mapping function $Act(a,ICP)$ which maps an integer $a$ to the corresponding action behavior on $ICP$. We can represent it as follows:
\begin{displaymath}
Act(a,ICP)=
\begin{cases}
Swap(T_{l},T_{r})  & a\in [1, I_{s}] \\
Override(O_{i}, Op_{j})  &a\in [I_{s}+1,  I_{s}+I_{o}] 
\end{cases}
\end{displaymath}
. When $Swap(T_{l},T_{r})$ is called, $l$ and $r$ satisfy the following conditions:
\begin{displaymath}
B_{l} \le a < B_{l+1}, r = a-B_{l} +2
\end{displaymath}
\begin{displaymath}
B_{k} = 
\begin{cases}
 1 &k= 1\\
 1 + \sum_{i = 2}^{k}(n-i + 1) &k\ge2
\end{cases}
\end{displaymath}
.When $Override(O_{i}, Op_{j})$ is called,  $i$ and $j$ satisfy the following conditions:
\begin{displaymath}
\begin{matrix}
  i=\lceil \frac{I_{s}+I_{o}+1-a }{\left | Op \right |}\rceil \\
  j=(I_{s}+I_{o}-a \mod \left | Op \right |) + 1
\end{matrix}
\end{displaymath}
. However, each query may have different action space restrictions, depending on the number of tables in the query graph. For example, in the query shown in Fig. \ref{fig:plan-step}, $Swap(T_{1}, T_{5})$ is considered an illegal action. To address the problem, we perform a validity check on the action space before the agent selects an action. We use $actionmask$ to prohibit illegal actions. To further prune the action space, we employ a heuristic rule. After executing $Swap(T_{l}, T_{r})$, the subsequent action can only be $Override(O_{i}, Op_{j})$, where $O_i$ is the parent node of $T_l$ or $T_r$. This also be implemented using the $actionmask$.\\
\textbf{Reward.} The design of the reward function plays a crucial role in DRL. It evaluates the effect of actions chosen by the agent. In planner, the reward $Reward_{t}^{e}$ at each step $t$ of one episode $e$ is composed of the sum of $Bounty_t^e$ and $Penalty_t^e$. The former provides positive feedback for actions selected by the agent, while the latter penalizes inappropriate actions.\\
% During the training process of planner, we maintain both the known best plan $CP^q_b$ for each query and the best plan $CP^p_b$ perceived by FOSS. And at each step $t$ of one episode $e$, a new incomplete plan $ICP_{t}^{e}$ and a new complete plan $CP_{t}^{e}$ will be generated. We determine reward based on these plans. The reward at each step $Reward_{t}^{e}$ is the sum of bounty $Bounty_t^e$ and penalty $Penalty_t^e$. And the reward for each episode is the sum of reward obtained at each step.\\
\textit{\underline{Bounty}.} First, we define the initial advantage function $Adv_{init}$ to calculate the initial advantage between two plans, indicating how much better $CP_r$ is compared to $CP_l$. 
\begin{displaymath}
    Adv_{init}(CP_{l},CP_{r})=\frac{U(CP_{l})-U(CP_{r})}{U(CP_{l})} \in (-\infty,1)
\end{displaymath}
, where $U(CP)$ is related to the performance of $CP$ (e.g., latency). However, fine-grained value is unsuitable for dynamically changing system workloads. We discretize initial advantage into several intervals. We take an $l$-element finite ordered point set $\{d_i |i\in \{1,2,...,l\},0\le d_1<...<d_l<1\}$ and use it to partition the interval $(-\infty,1]$ into $l+1$ subintervals, resulting in the set of ordered subintervals $D = \{(d_i, d_{i+1}]|i \in\{0,1,...,l\},d_0\to-\infty,d_{l+1}=1\}$. We use $D_k$ to represent the $k$-th element in $D$ and set $\hat{D}_k  = (d_k + d_{k-1})/2$, $\hat{D}_0=0$. We define the final advantage function $Adv$ that takes a pair of plans as input and outputs a score. 
\begin{equation}\label{advantage function}Adv(CP_{l},CP_{r}) = k-1 \ \text{ if }  Adv_{init}(CP_{l},CP_{r})\in D_k  
\end{equation}
For each episode $e$ of query $Q$, the planner starts from the original plan $CP^Q_{ORI}$ and aims to generate a better plan $CP_{t}^{e}$ than the plans from previous $t-1$ steps at step $t$. We assign the step-bounty $pb_t^e = Adv(\overline{CP}^e_{t-1},CP_{t}^{e})$, where $\overline{CP}^e_{t-1}$ is the estimated optimal plan among all plans from the previous $t-1$ steps. Naturally, when an episode $e$ concludes, the final output plan $\overline{CP}^e$ generated by FOSS (i.e., the plan to be executed) should carry significant weight in the reward. Therefore, we add the episode-bounty $eb^e$ based to the performance of $\overline{CP}^e$. To comprehensively evaluate the performance of $\overline{CP}^e$, we select the best-performing plan and the median-performing plan from the set of executed plans for query Q that outperform the original plan. These plans, along with the original plan, form the reference plan set $CP^{ref}$. Then we calculate the actual bounty of each reference plan using the formula $refb_i = Adv_{init}(CP^{Q}_{ORI},CP^{ref}_{i})$ to serve as reference bounty. Finally, we estimate the $adv_i=Adv(CP^{ref}_i,\overline{CP}^e)$ to determine the ranking range of $\overline{CP}^e$. The $eb^{e}$ is defined as follows:
\begin{displaymath}
eb^e=\sum_{i=1}^{3} (\hat{D}_{adv_{i}}+ \frac{adv_{i}}{l})\times (refb_{i-1}-refb_{i})
\end{displaymath}
, where $refb_0$ is set to 1, indicating the upper limit value. $Bounty_t^e$ is represented as follows:
\begin{displaymath}
    Bounty_{t}^{e} = pb_t^e + \eta \times \lfloor\frac{t}{maxsteps}\rfloor \times eb^e
\end{displaymath}
, where $\eta$ is a constant representing the weight of episode-bounty relative to step-bounty.\\
\textit{\underline{Penalty}. }To encourage the agent to reach each different state with as few steps as possible, we implement a penalty measure. In planner, reaching a certain state from the initial state can involve multiple action sequences of varying lengths, where an action sequence refers to a combination of $Swap$ and $Override$ operators. In order to gain more rewards, the agent may opt for inappropriate action sequences. For example, assume that the execution latency for the hash join operator at position $O_{i}$ is $L_{1}$, $L_{2}$ for merge join, and $L_{3}$ for nested loop join. Suppose $L_{1}>L_{2}>L_{3}$, and the original plan choose using hash join. In order to obtain more bounty, the agent may choose to initially override the method at $O_{i}$ to a merge join, and then in subsequent steps, override it to a nested loop join. To address such issue, we calculate the minimum number of steps required from the original incomplete plan to the current incomplete plan, denoted as $minsteps(ICP_{t}^{e})$. Additionally, we introduce a penalty coefficient $\gamma > 0$ to represent the weight of penalty in reward. We set the penalty value according to the following formula: 
\begin{equation}
Penalty_{t}^{e}= \gamma \times (minsteps(ICP_{t}^{e})-t)
\end{equation}
. If the current step  is already the minimum, the penalty value will be 0. Otherwise, it will have a negative impact on the reward. It ensures that the agent can learn useful knowledge more effectively.

In FOSS, $\eta$ is set to 12 and $\gamma$ is set to 2.\\
\textbf{Agent.} In planner, the agent mainly consists of two models: the state network $\phi$ and the action selector $\pi$. The former is a transformer-based network used to process the state. It takes the state $s$ as input and outputs the state representation vector $statevec$. Further details about the state network will be explained in \ref{state repre}. We use multi-layer fully connected neural network to serve as the action selector. It takes the $statevec$ and the $actionmask$ as input and outputs the action encoding $a$ that maximizes the cumulative expected reward.
\begin{displaymath}
    \phi(s) \to statevec, \pi(statevec,actionmask) \to a
\end{displaymath}
\textbf{Environment.} Its main functions include: \romannumeral1) providing new state based on the action taken by the agent; \romannumeral2) assessing the agent's action to provide reward.
For \romannumeral1), we use the DBMS optimizer $\Gamma_p$ to provide new state. We can express it using the following formula:
\begin{displaymath}
\begin{cases}
\Gamma_p(Q,/) \to CP_t  &  t=0\\
\Gamma_p(Q,ICP_{t}) \to CP_t  & t>0
\end{cases}
\end{displaymath}
. In the case of the initial step, it takes the query $Q$ as input and outputs a complete plan $CP$. In non-initial steps, it takes both the incomplete plan $ICP$ and the query $Q$ as input, generating a complete plan $CP$ steered by $ICP$. We then encode the $CP$ to obtain the new state by using \eqref{state}. 
For \romannumeral2), after obtaining $CP$, we provide a reward as described in the part of \textit{Reward}. The key difference is that in the real environment, we use the DBMS executor $\Psi_p$ to execute the plan and then use the latency to represent the performance of plan $U(CP)$. Then we calculate the advantage and the final reward. In the simulated environment, the advantage value is evaluated by asymmetric advantage model without executing the plan. The specific details will be discussed in \ref{Advantage model}.

\begin{algorithm}
	\caption{Planner Training Process}
	\label{alg:planner}
	\begin{algorithmic}[1] 
		% \Require Workload $W$.
		% \Ensure  CandidatePlanSet $\{CP_{t}|t=0,1,2,...,maxsteps\}$. 
        \For {Query $Q$ in Workload $W$}
        \State $ CP^e_{0} (CP^Q_{ORI}) \gets \Gamma_p(Q,/)$
        \State $ICP^e_{0} \gets Extract(CP^e_{0})$
        \State Episode Buffer Set $T = [ICP^e_{0}]$
        \State Estimated Optimal Plan $\overline{CP}^e_{0} \gets CP^e_{0}$
		\For {$t=1 \to maxsteps$}
        \State $actionmask \gets ICP^e_{t-1}$
        \If {$t>1 \cap  a^e_{t-1} \in [1, I_{s}]$}
            \State $actionmask \gets LimitSpace(actionmask)$
        \EndIf
        \State  $s^e_{t-1} = State(CP^e_{t-1})$
		\State  $statevec^e_{t-1} = \phi(s^e_{t-1})$
        \State  $a^e_{t} = \pi(statevec^e_{t-1},actionmask)$
        \State  $ICP^e_{t} = Act(a^e_{t},ICP^e_{t-1})$
        \State  $CP^e_{t} \gets \Gamma_p(Q,ICP^e_{t})$
        \State $r^e_{t}= Penalty^e_t$
        \If {$ICP^e_{t} \notin T$}
            \State  $r^e_{t}=r^e_{t}+Bounty^e_t $ 
             \State $Add\ ICP^e_{t}\ to\ T$
        \EndIf
        \State $\overline{CP}^e_t\gets CP^e_t \text{\   \  if\ \  } Adv(\overline{CP}^e_{t-1}, CP^e_{t}) > 0$
        \State  $experiences \gets Collect(s^e_{t-1},a^e_{t},r^e_{t},s^e_t)$
		\EndFor
  \EndFor
    \State $UpdateAgent(experiences)$
        % \State \Return $\{CP_{t}|t=0,1,2,...,maxsteps\}$. 
	\end{algorithmic}
\end{algorithm}
Algorithm \ref{alg:planner} shows the overall training process of planner. When a query $Q$ is inputted, the DBMS optimizer generates the original plan $CP^{Q}_{ORI}$ (Line 2). From $CP^{Q}_{ORI}$, the planner extracts the $ICP^e_{0}$ composed of the join order and join methods (Line 3). The planner initializes the episode buffer set $T$ and the estimated optimal plan $\overline{CP}^e_{t}$ (Line 4-5). For every step, the planner calculates the forbidden actions based on the $ICP^e_{t-1}$ (Line 7). If the previous action is $Swap$ operator, the current action will be restricted to only overriding the adjacent join method (Line 8-10). Next, the planner represents the previous state based on the $CP^e_{t-1}$ and step status (Line 11-12) and evaluates the action based on the state representation $statevec^e_{t-1}$ and $actionmask$ (Line 13). Then it applies this action to $ICP^e_{t-1}$ and produces a new incomplete plan $ICP^e_{t}$ (Line 14). The $ICP^e_{t}$ will steer the DBMS optimizer to output a complete plan $CP^e_{t}$ (Line 15). The planner computes the penalty value and uses it to initialize the reward (Line 16). If the current incomplete plan $ICP^e_{t}$ has not appeared in this episode, we add the bounty to reward and add the $ICP^e_t$ to $T$ (Line 17-20). If the current plan is better than $\overline{CP}^e_{t-1}$, the planner updates the $\overline{CP}^e_{t}$ with $CP^e_t$ (Line 21). Experiences are collected in order to update the agent (Line 22). Finally, upon concluding the exploration of the current workload, the agent is updated using $experiences$ (Line 25).

\section{Asymmetric Advantage Model}\label{rewardmodel}
In this section, we will provide a detailed description of the AAM (i.e., asymmetric advantage model). We begin by introducing the state representation, followed by the advantage model. Finally, we will discuss the design of the loss function.
\subsection{State Representation} \label{state repre}
As mentioned in \eqref{state}, state consists of both plan encoding and step encoding. We have introduced step encoding. Next, we will provide a detailed explanation of how we encode the complete plan. Finally, we will introduce the state network which transforms state to state representation.\\
\textbf{Plan Encoding. }Our work is structured based on the QueryFormer \cite{QueryFormer}, which is a tree transformer model for query plan representation. Following their work, we first extract node features, including operators, predicates, joins, and tables. But we abstain from employing histogram data and sampling information due to computational efficiency concerns. To capture the node structural features of the query plan, we encode the height of nodes, which is defined as the length of the longest downward path from the node to a leaf node in the plan tree. Additionally, different from \cite{QueryFormer}, we encode the structure type of each node. We classify nodes into four types: left nodes, right nodes, no-siblings nodes, and root nodes. We use the labels 0, 1, 2, and 3 to represent these four types of nodes, obtaining the node structure feature. To capture the tree structural features of the plan, we utilize attention network with attention mask. As the original attention network will attend to all nodes in the plan tree, which is not entirely reasonable, we aim to focus the attention network on meaningful structural information. To achieve this, we mask the mutual influence between unreachable nodes in the query plan tree, setting the attention score to 0 between two unreachable nodes and 1 between two reachable nodes. It differs from their work, which uses biased weights to adjust attention scores among nodes at different levels. We argue that assuming the same level of difference affects attention scores similarly is not reasonable for different nodes in various query plans or within the same query plan. This limitation hinders the model's power.\\
\textbf{State Network. }We individually pass the extracted four types of node features through specific embedding layers to obtain the vector representation of each feature. Subsequently, we concatenate the four vectors to obtain the node representation, denoted as $N_i$, for the $i$-th node. We use dedicated embedding layers to represent the height and node structure features, obtaining $height_i$ and $ns_i$ for each node. Finally, we concatenate $N_i$, $height_i$, $ns_i$ and feed them into the multi-head attention network with attention scores. This process ultimately yields the representation of the query plan. Subsequently, it is concatenated with the step encoding and passed through a linear layer to generate the final state representation vector $statevec$.
\subsection{Advantage Model } \label{Advantage model}
As described in \eqref{advantage function}, we need to calculate the advantage between two plans. In the real environment, precise advantage values are obtained by executing the plans offline. In simulated environment, we estimate the advantage values using an advantage model which takes plan pairs $(CP_{l}, CP_{r})$ as input. It initially encodes the two input plans separately, transforming them into $(statevec_l, statevec_r)$ using the state network. Next, it incorporates position information for each $statevec$ to obtain $(statevec_l\oplus pos_{left}, statevec_r \oplus pos_{right})$. Finally, the pair is passed to the final output layer, where it's mapped to the scores (i.e., the output of $Adv$). The AAM $\theta_{adv}$ can be described as follows: 
\begin{displaymath}
\begin{split}
\theta_{adv}(CP_l,CP_r)&\to FC^2(FC^1(\phi(State(CP_l))\oplus pos_{left})\\
&-FC^1(\phi(State(CP_r))\oplus pos_{right}))
\end{split}
\end{displaymath}
, where $FC$ represents fully-connected network. In the simulated environment, $\theta_{adv}$ serves as the advantage function $Adv$, and training data in the format $\{(CP_l, CP_r), Adv(CP_l, CP_r)\}$ will be used to update $\theta_{adv}$.

Considering that the dynamic changes in system workload may cause the labels of plan pairs with similar execution latency to change continuously and to select significantly superior plans, we take the point set $\{0.05, 0.50\}$ to divide the interval $(-\infty, 1]$, resulting in three scores (i.e., $\{0,1,2\}$) for the output of the $Adv$.

\subsection{Asymmetric Loss}
Since plans obtained from traditional optimizer are often of reasonable performance, starting with such plans as baselines and making modifications often yields more inferior plans. As a result, training samples for $\theta_{adv}$ are skewed towards a larger number of samples with a score of 0. This uneven distribution of samples across different labels will adversely affect the training of the $\theta_{adv}$. Inspired by Asymmetric Loss \cite{ASL}, FOSS addresses the issue by increasing the weights of difficult-to-classify samples. Additionally, to prevent overfitting during training, FOSS also employs label smoothing.

For a given training samples set $\{x_i=(CP_l^i, CP_r^i), y_i=Adv(CP^i_l, CP^i_r)|i=1,2,...,N\}$, where $N$ represents the number of training samples, we employ one-hot encoding $h_{i,j}$ to represent the true label $y_{i}$:
\begin{displaymath}
    h_{i,j} =
\begin{cases}
1, & j =y_i \\
0, & j\ne y_i
\end{cases}
\end{displaymath}
. We use $Softmax$ to convert the output of $\theta_{adv}$ into score probability distribution. $p_{i,j}$ represents the probability of the $j$th label in the $i$th sample, and $\hat{p}_{i,j}$ indicates the sample's classification difficulty. Smaller $\hat{p}_{i,j}$ values suggest higher difficulty, and vice versa.
% . We use the $Softmax$ function to map the output of $\theta{adv}$ into the probability distribution of the scores. The probability of the $jth$ label in the $ith$ samples is denoted as $p_{i,j}$. We use $\hat{p_{i,j}}$ to represent the difficulty of classifying the samples, as defined in \eqref{phat}. A smaller $\hat{p_{i,j}}$ value indicates a more difficult-to-classify samples, and vice versa.
\begin{align}\label{phat}
\hat{p}_{i,j}=
\begin{cases}
 p_{i,j}    & h_{i,j}=1\\
 1-p_{i,j} & h_{i,j}=0
\end{cases}
\end{align}
To make $\theta_{adv}$ pay more attention to difficult-to-classify samples, FOSS introduces a decay factor of $(1-\hat{p}_{i,j})^\gamma$, where $\gamma$ represents the decay coefficient. To emphasize the contribution of positive samples, we assign different decay coefficient to positive and negative samples. We denote the decay coefficient for positive samples as $\gamma_+$ and the decay coefficient for negative samples as $\gamma_-$ and set $\gamma_+ < \gamma_-$. This is represented by the following formula: 
\begin{displaymath}
\begin{split}
L_{i,j}=&-(1-\hat{p}_{i,j})^\gamma log(p_{i,j})\\=&\begin{cases}
    (1-p_{i,j})^{\gamma+}log(p_{i,j}) &h_{i,j}=1\\
    p_{i,j}^{\gamma-}log(p_{i,j}) & h_{i,j}=0
\end{cases}
\end{split}
\end{displaymath}
. We modify the probability distribution as follows:
\begin{displaymath}
    \hat{h}_{i,j} =
\begin{cases}
1-\epsilon, & j =y_i\\
\frac{\epsilon}{K-1}, & j\ne y_i
\end{cases}
\end{displaymath}
, where $\epsilon$ is a hyperparameter and $K$ represents the number of scores. In FOSS, $K$ is set to 3 and $\epsilon$ is set to 0.1. The final loss function can be expressed as follows:
\begin{displaymath}
    LOSS=-\sum_{i=1}^{N}\sum_{j=1}^{K} \hat{h}_{i,j}L_{i,j}.
\end{displaymath}
\section{Simulated Learner}
In this section, we will begin by introducing the simulated environment. Next, we will provide an explanation of how FOSS explores the real environment and trains the agent with the simulated environment.
\subsection{Simulated Environment}\label{environment}
As described in Algorithm \ref{alg:planner}, data in the form of $\{s, a, r, s^{'}\}$ are collected to update the agent. However, in many environments, obtaining the real state and reward is challenging due to high costs or security concerns. To overcome this challenge, researchers propose simulating environment using neural networks to accelerate agent's learning. A simulated environment always includes State Transition Dynamics Model $\theta_{p}(s,a) \to s^{'}$ and Reward Function Model $\theta_{r}(s,a) \to r$ \cite{surveymbrl}. They will have the agent interact with the real environment and use the real experiences to perform supervised training of the simulated environment \cite{surveymbrl2020}. Then the agent efficiently interacts with the simulated environment to generate simulated experiences, which are used to update the agent accordingly.

In the context of planner, state is consist of $CP$ and step status, while step status is determined and $CP$ can be obtained from DBMS optimizer with low cost, especially within the guidance of $ICP$. Therefore, we can assign $\theta_{p}(s,a)$ with DBMS optimizer $\Gamma_p$. However, obtaining the true reward requires executing the complete plan, which can be time-consuming. We only need to be aware of how we can simulate $\theta_{r}(s,a)$ to provide rewards feedback timely. Using the AAM $\theta_{adv}$ that we have established to assign rewards is a natural idea. Therefore, we employ the DBMS optimizer $\Gamma_p$ and $\theta_{adv}$ to construct a simulated environment $\hat{E}(\Gamma_p,\theta_{adv})$.
\subsection{Training Loop}
\begin{figure}[tb]
  \centering
  \includegraphics[width=\linewidth]{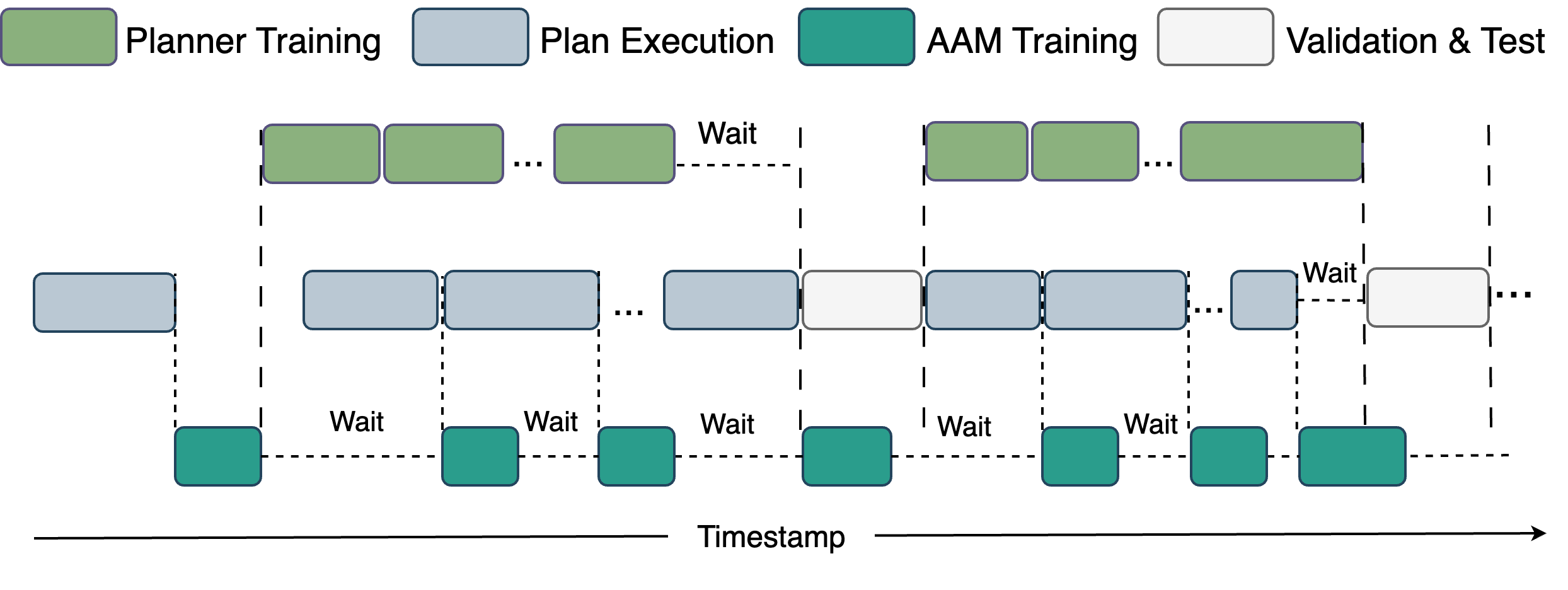}
  \caption{The training loop of FOSS.}
  \label{fig:trainingloop}
\end{figure}
In this subsection, we will discuss how to train the agent of the planner effectively using the simulated environment and the real environment, and how to collect training data for updating AAM.

As depicted in Fig. \ref{fig:trainingloop}, FOSS concurrently conducts training for the planner, AAM, and plan execution. It begins by randomly initializing the planner, which then generates candidate plans for sampled queries from the workload. After executing these plans, FOSS collects their execution latency into the execution buffer. Then, as described in \ref{Advantage model}, we utilize the data in execution buffer to form the training data for AAM and update it. Once AAM training concludes, the planner interacts with the simulated environment $\hat{E}(\Gamma_p,\theta_{adv})$, following Algorithm \ref{alg:planner}. Simultaneously, FOSS gathers promising plans to be executed based on the feedback of AAM during planner training. FOSS also periodically samples queries from the workload at random and collects candidate plans to be executed. The listener process triggers plan execution by the executor upon receiving new plans to be executed and aggregates execution latency into the execution buffer. Upon accumulating a sufficient number of executed plans, FOSS reorganizes the training data from the execution buffer and retrains the AAM. During plan execution, we implement dynamic timeouts to prevent poorly performing plans from significantly hindering the training progress. The timeout duration is set to 1.5 times the execution latency of the original plan. If the execution latency of a newly generated plan exceeds this threshold, it is terminated and labeled as a timeout. When organizing the AAM training data, we filter out plan pairs that involve timeouts for both plans. While this timeout strategy may lead to a reduction in the number of training samples for AAM and potentially lower the prediction accuracy of AAM, it significantly increases the rate of data collection and improves the training efficiency of FOSS.

\section{Experiments}\label{experiments}
In this section, we outline our experimental configurations and analyze the performance of FOSS while also conducting comparisons with SOTA methods. Finally, we analyze the effects of various components of FOSS and demonstrate the additional features of FOSS.
% Finally, regarding the functions and roles of various components  of FOSS, we will address the following questions:
% \begin{itemize}
% \item Starting from the original plan, how many steps does FOSS typically require to obtain an excellent query plan?
%     \item What effect does the use of a simulated environment bring about on FOSS?
%     \item What effect does the use of multi-agent strategy bring about on FOSS?
% \end{itemize}

\subsection{Experiment Setup}
\noindent\textbf{Experiment Support.} We deploy FOSS alongside the SOTA methods on an Ubuntu 18.04 system, featuring a hardware configuration that includes Intel(R) Xeon(R) Gold 5118 CPU @2.30GHz, 256GB of memory, and a Geforce RTX 3090 GPU. We implement FOSS using PyTorch and utilize Ray \cite{Ray} for the RL component. We utilize PPO \cite{PPO} as the base RL algorithm due to its effectiveness in mitigating differences in the action distribution before and after agent updates through KL divergence. It further ensures the accuracy of the estimated reward provided by AAM in the simulated environment.\\
\textbf{Workloads.} We evaluate the performance of all the optimizers on three different benchmarks: Join Order Benchmark (JOB) \cite{JOB}, TPC-DS\cite{tpcds}, and Stack \cite{Bao}.\\
\underline{JOB} builds on the real-world IMDb dataset, which consists of 21 relations with a total data size of 3.6GB. Within the JOB workload, there are 33 query templates, encompassing 113 individual queries. We follow Balsa's \cite{Balsa} random partitioning, splitting queries into 94 for training and 19 for testing.\\
\underline{TPC-DS} is a standard benchmark. We utilize its data generation tool to create a dataset of 10GB in size. It includes 99 query templates. Nevertheless, a majority of these templates do not meet the requirements of some SOTA methods (such as requiring select-project-join queries) and the left-deep constraint of FOSS. We eventually opt for 19 query templates\footnote{The selected template numbers for TPC-DS are 3, 7, 12, 18, 20, 26, 27, 37, 42, 43, 50, 52, 55, 62, 82, 91, 96, 98 and 99.}. From each template, we generate 6 individual queries, and from these, we randomly select 5 queries for the training workload and the remaining 1 query for the testing workload.\\
\underline{Stack} comprises 18 million questions and answers collected from StackExchange websites such as StackOverflow.com over a decade. It is generated by \cite{Bao} and occupy a total space of 100GB. Within the workload, there are 16 query templates. Similar to TPC-DS, we filter out the templates that do not meet the requirements and then randomly select 10 queries from the remaining 12 templates\footnote{The selected template numbers for Stack 1, 4, 5, 6, 7, 8, 11, 12, 13, 14, 15, 16.}. For each template, 8 queries are assigned for the training workload, and 2 queries for testing.\\
\textbf{Comparision.} We employ PostgreSQL as baseline and compare FOSS with the SOTA methods, including the two plan-steerer methods (i.e., Bao \cite{Bao} and HybridQO \cite{HybridQO}), and the two plan-constructor methods (i.e., Balsa \cite{Balsa} and Loger \cite{LOGGER}).\\
\textit{\underline{PostgreSQL}} is an open-source DBMS with a traditional optimizer. We employ it as a representative of traditional optimizer.\\
\textit{\underline{Bao}} steers traditional optimizer to select better plan by employing a value model to choose suitable hint set. By default configuration, it has 5 hint sets available. We configure it according to the instructions \cite{BaoProject}. \\
\textit{\underline{HybridQO}} combines learned-based optimizer with traditional optimizer. It employs Monte Carlo Tree Search to discover promising leading join order and uses these as hints to steer traditional optimizer to generate better plans. We replicate their experimental setup as described in \cite{HyperQOproject}.\\
\textit{\underline{Balsa}} is an end-to-end query optimizer that operates independently from traditional optimizer. We follow the instructions outlined in its source code \cite{balsaproject} to deploy it.\\
\textit{\underline{Loger}} focuses on join order and join method optimization, which aligns with FOSS. It learns to join individual tables in a bottom-up manner but restricts specific join methods instead of directly selecting one for each join. We implement it according to its source code \cite{logerproject}.\\
\textbf{Expert Engine.} For a fair comparison, We conduct all experiments using PostgreSQL 12.1. We set up PostgreSQL with 32GB shared buffers and disable GEQO due to the usage of \textit{pg\_hint\_plan}. We employ it as the expert engine in our experiments.\\
\textbf{Metrics.} We deploy all learned query optimizers on the training workload and evaluate their performance on the testing workload using the two metrics: \romannumeral1) Geometric Mean Relevant Latency (GMRL): it is introduced in \cite{Tree-lstm} and demonstrates the optimization effectiveness at the query level. It can be expressed using the formula $GMRL = \sqrt[|W|]{\prod_{q\in W}\frac{ET_{l}^q}{ET_{e}^q}}$, where $ET_{l}$ and $ET_{e}$ respectively represent the execution latency of the learned-based optimizer and the expert optimizer. \romannumeral2) Workload Relevant Latency (WRL): it primarily reflects the overall latency level of the entire workload, as previously employed in the approach \cite{Lero,Balsa,LOGGER}. It can be expressed using $WRL = \frac{\sum_{q}^{W} (ET^q_{l}+OT^q_{l})}{\sum_{q}^{W}(ET^q_{e}+OT^q_{e})}$, where $OT_{l}$ and $OT_{e}$ respectively represent the optimization time of the learned-based optimizer and the expert optimizer. For the two metrics, a smaller value indicates better performance. A value greater than 1 implies inferiority compared to the expert optimizer, while a value less than 1 signifies the opposite.
% Its computation involves the ratio of the total latency of the entire workload executed by the learned-based optimizer to the total latency observed in expert optimizer.
\subsection{FOSS Performance and Comparison}
Unless specified otherwise, we set the $maxsteps$ to 3 and use 900 episodes for each update of agent during the training phase of planner. We deploy all methods on each benchmark, running them three times with different random seeds.
\begin{figure}[tb]
   \centering
  \includegraphics[width=\linewidth]{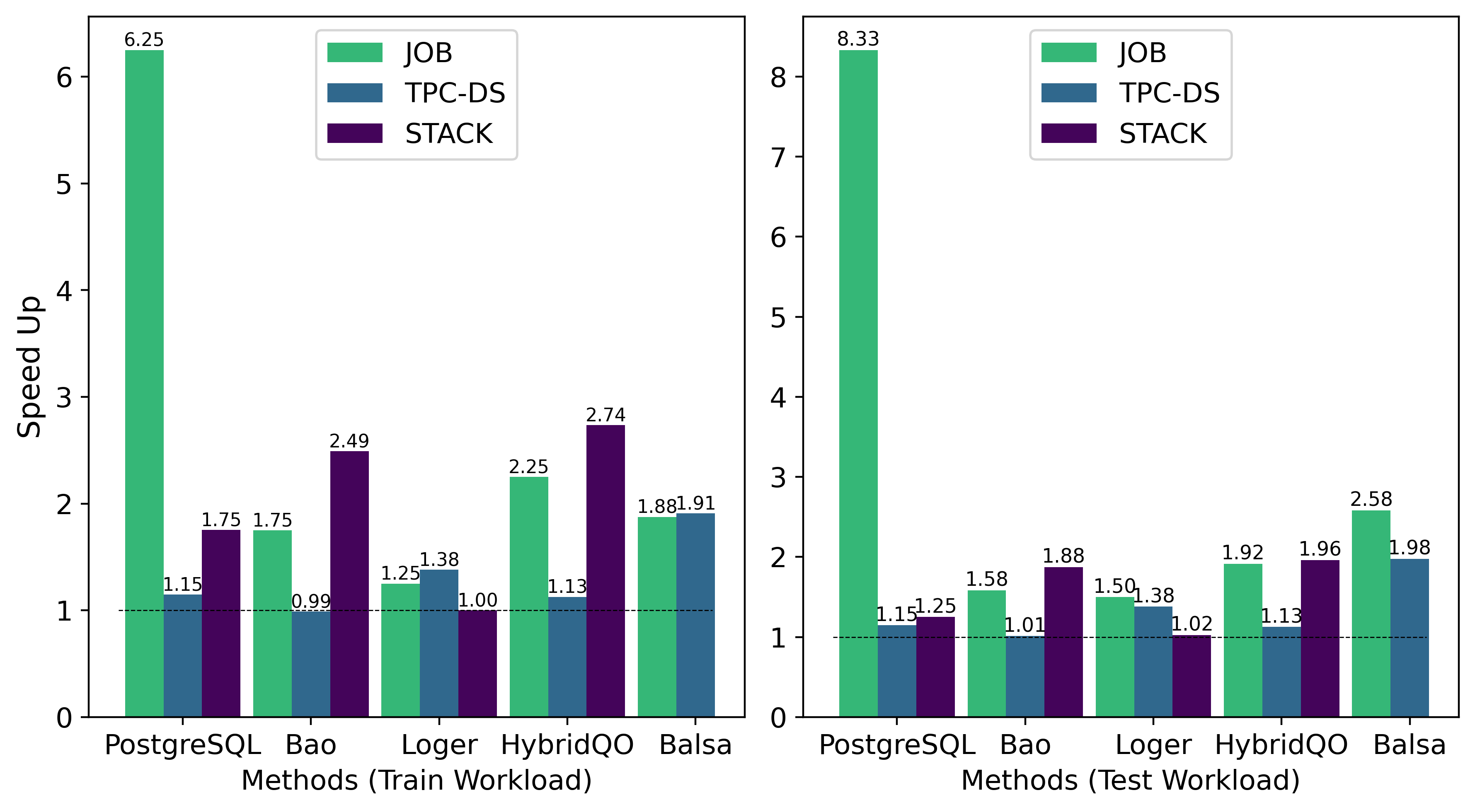}
%   \subfigure[Training workload]
% {
%     \begin{minipage}[b]{.45\linewidth}
%         \centering
%         \includegraphics[scale = 0.44]{figures/speeduptrain.png}
%         % \caption{Optimization time.}
%     \end{minipage}
% }
%  \subfigure[Testing workload]
% {
%     \begin{minipage}[b]{.45\linewidth}
%         \centering
%         \includegraphics[scale = 0.44]{figures/speeduptest.png}
%     \end{minipage}
% }
\caption{The relative speedup of FOSS compared to other methods across various workloads.}
\label{fig:speedup}
\end{figure}
\subsubsection{Performance Overview}
\begin{table*}[tb]% h asks to places the floating element [h]ere.
\centering
\caption{The performance of various methods and training time on different workloads.}
  \label{tab:performance}
  \tabcolsep=0.05cm
    \begin{tabular}{c|ccc|ccc|ccc|ccc|ccc}
    % \cline{2-7}
    \hline
    % \  &Performance (WRL/GMRL)\\
     \multirow{2}{*}{Methods}& \multicolumn{3}{c|}{WRL / train}&\multicolumn{3}{c|}{GMRL / train} &  \multicolumn{3}{c|}{WRL / test}& \multicolumn{3}{c|}{GMRL / test}& \multicolumn{3}{c}{Workload Runtime (s)}\\ 
     \cline{2-16}
      & JOB  & TPC-DS& Stack & JOB  & TPC-DS& Stack   & JOB  & TPC-DS& Stack & JOB  & TPC-DS& Stack & JOB  & TPC-DS&Stack \\
     \hline
    PostgreSQL& 1.00&  1.00&1.00& 1.00& 1.00&1.00& 1.00& 1.00& 1.00& 1.00& 1.00& 1.00& 161.50& 35.59&36.31\\
    Bao & 0.28&  \textbf{0.86}&1.42& 0.70&  0.92&0.81& 0.19& 0.88& 1.50& 0.71& 0.96& 0.95& 30.49& 31.32&54.47\\
    Balsa & 0.30&  1.66&TLE& 1.30& 2.54&TLE& 0.31& 1.72& TLE& 1.69& 2.59& TLE& 49.76& 61.22&TLE\\
    Loger& 0.20& 1.20& \textbf{0.57}& 0.60& 1.06& 0.64& 0.18& 1.22& 0.82& 0.64& 1.09& 0.77& 29.07& 42.71&29.77\\
    HybridQO& 0.36&  0.98&1.56& 0.65&  0.95&0.92& 0.23& 0.98& 1.57& 0.74& 1.04& 0.97& 36.92& 34.88&57.01\\
    FOSS & \textbf{0.16}&   0.87&\textbf{0.57}& \textbf{0.48}&  \textbf{0.90}&\textbf{0.63}& \textbf{0.12}& \textbf{0.87}& \textbf{0.80}& \textbf{0.61}& \textbf{0.91}& \textbf{0.72}& \textbf{19.38}& \textbf{30.96}&\textbf{29.05}\\
   \hline
\end{tabular}
\end{table*}

\begin{figure*}[htbp]
\centering
\includegraphics[width=\linewidth]{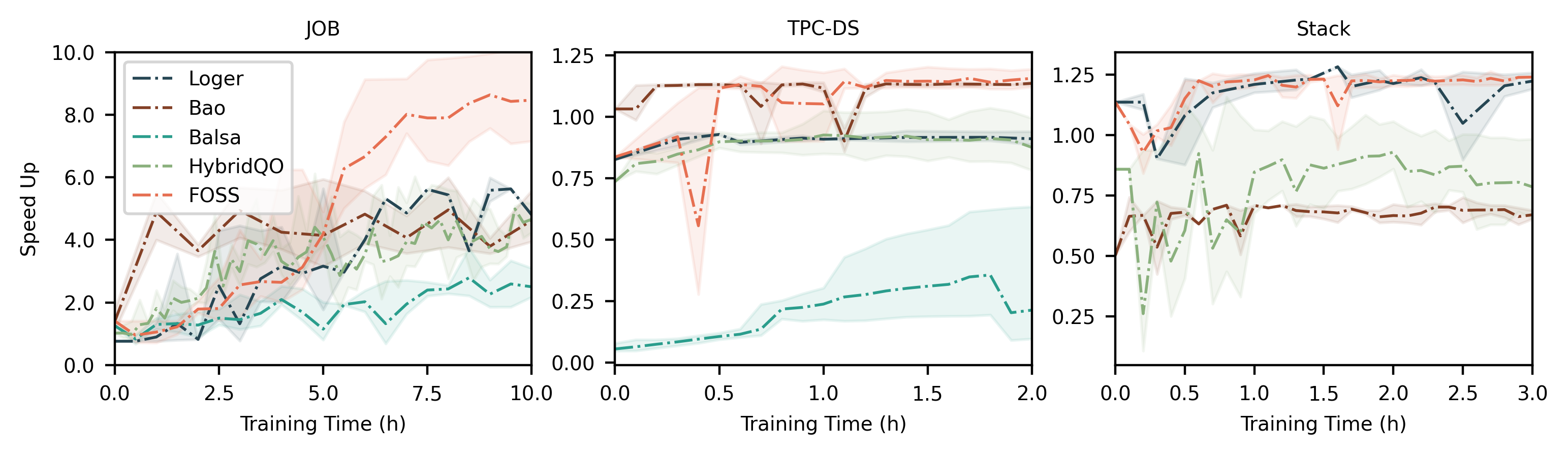}
% \subfigure[JOB]
% {
%     \begin{minipage}[b]{.32\linewidth}
%         \centering
%         \includegraphics[scale = 0.79]{figures/JOBRand.png}
%     \end{minipage}
% }
% \subfigure[TPC-DS]
% {
%  	\begin{minipage}[b]{.31\linewidth}
%         \centering
%         \includegraphics[scale = 0.79]{figures/TPCDS.png}
%     \end{minipage}
% }
% \subfigure[STACK]
% {
%  	\begin{minipage}[b]{.31\linewidth}
%         \centering
%         \includegraphics[scale = 0.79]{figures/STACK.png}
%     \end{minipage}
% }
% \subfigure[GMRL of JOB]
% {
%     \begin{minipage}[b]{.3\linewidth}
%         \centering
%         \includegraphics[scale = 0.58]{figures/JBRGMRL.png}
%     \end{minipage}
% }
% \subfigure[GMRL of TPC-DS]
% {
%  	\begin{minipage}[b]{.3\linewidth}
%         \centering
%         \includegraphics[scale = 0.58]{figures/DSGMRL.png}
%     \end{minipage}
% }
% \subfigure[GMRL of STACK]
% {
%  	\begin{minipage}[b]{.3\linewidth}
%         \centering
%         \includegraphics[scale = 0.58]{figures/STGMRL.png}
%     \end{minipage}
% }
\caption{Training curves of speedup relative to expert optimizer.}
\label{fig:FOSSPerformance}
\end{figure*}
In this subsection, we compare the overall performance of FOSS with the other optimizers. We ensure that all SOTA methods are evaluated after convergence. Table \ref{tab:performance} shows the results of each method on various workloads. Notably, Balsa's training on Stack is hindered by catastrophic plans generated during the initial phase, preventing it from completing within a reasonable timeframe. Fig. \ref{fig:speedup} illustrates the relative speedup of FOSS compared to other optimizers across various workloads.

Overall, FOSS outperforms the expert optimizer (i.e., PostgreSQL optimizer) with speedup of $6.25\times$, $1.15\times$, and $1.75\times$ in total latency for the training workload (JOB, TPC-DS, Stack) and $8.33\times$, $1.15\times$, and $1.25\times$ for the testing workload, respectively. Compared to other learned optimizers, FOSS consistently outperforms the other four methods in terms of average performance for both training and testing workloads. FOSS outperforms Bao, HybridQO, Loger and Balsa with average speedup of $1.74\times$, $2.04\times$, $1.21\times$ and $1.90\times$ on training workloads, and $1.49\times$, $1.67\times$, $1.33\times$ and $2.28\times$ on testing workloads, respectively.

Specially, the JOB workload entails complex query statements and non-uniform data distributions, posing a significant challenge to query optimizers. Therefore, the performance on the JOB workload is more indicative of an optimizer's capabilities. FOSS significantly outperforms other SOTA methods on the JOB workload, demonstrating its superior performance.

As shown in Table \ref{tab:performance}, FOSS not only outperforms other methods in terms of total workload latency but also demonstrates superior performance in the GMRL metric. Compared to Balsa, FOSS benefits from the assurance provided by the original plans and efficiently interacts with the simulated environment, enabling more exploration. In contrast to Loger, Bao, and HybridQO, FOSS utilizes finer-grained optimization, increasing the likelihood of generating superior plans than original plans. These factors contribute to FOSS outperforming other methods in terms of the GMRL metric and surpassing the expert optimizer across a wider range of queries, highlighting its stronger optimization capabilities. This will be further confirmed in \ref{Knwon Best Plan}.
% As shown in Table \ref{tab:performance}, FOSS not only outperforms other methods at the workload total latency but also exhibits superior performance in GMRL metric. Compared to Balsa, FOSS can efficiently interact with the simulated environment, allowing for more exploration and guarantee by the orginal plan. Compared to Loger, Bao and HybridQO, FOSS employs finer-grained optimization, resulting in a higher likelihood of generating better plans. These factors enable FOSS to surpass the expert optimizer across a broader range of queries, indicating its stronger optimization capabilities. 
% Worth mentioning is the performance on the JOB workload, where Bao and Balsa exhibit a relatively small difference compared to FOSS in WRL, but a more significant gap emerges in GMRL. 
% However, these advantages come with an associated cost, leading to relatively longer training time on the JOB compared to HybridQO and Bao. 

% Nevertheless, in comparison to Balsa, another MDP-based method, FOSS exhibits a significant reduction in training time. While on STACK and TPC-DS, the training time difference among FOSS, Bao, and HybridQO is not pronounced and is much less than that of Balsa, FOSS surpasses all other methods, demonstrating the superiority of our approach.

\subsubsection{Training efficiency}
Fig. \ref{fig:FOSSPerformance} illustrates the training curves of the average speedup relative to the expert optimizer over time for different learned optimizers on the test workload. The shaded area indicates the range between the minimum and maximum values. It can be observed that benefiting from the assurance provided by the original plans, FOSS can swiftly demonstrate its optimization effects and surpass the performance of the expert optimizer. On the JOB workload, FOSS achieves superior performance compared to other SOTA methods after 5 hours of training and converges to the best performance level after around 8 hours. On the TPC-DS workload, FOSS converges in about 1 hour and outperforms Loger, Balsa, and HybridQO, reaching performance levels comparable to Bao. On the Stack workload, FOSS also converges in about 1 hour, exhibiting a training efficiency curve similar to Loger.
% Fig. \ref{fig:FOSSPerformance} illustrates the training curves of FOSS on JOB, TPC-DS, and STACK. The solid red line represents the average change in metrics over five runs on the training workload, while the blue line indicates the average change in metrics on the testing workload. The shaded area indicates the range between the minimum and maximum values. 

% The line at 1.0 indicates the performance of the expert optimizer. Benefiting from the assurance provided by the original plans, FOSS can swiftly demonstrate its optimization effects and surpass the performance of expert optimizer. 

% On the JOB workload, the complexity of queries and data distribution allows FOSS to have more room for optimization. After about 5 hours of training, FOSS achieves convergence, ultimately saving $64\%$ of the latency on the testing workload relative to expert optimizer. A steady decrease on the GMRL curve until convergence to $0.64$ on the testing workload demonstrates FOSS's ability to optimize the majority of original plans and achieve high performance. On the training workload, FOSS shows a similar convergence trend, indicating its good generalization. 

% On the TPC-DS and STACK workloads, FOSS occasionally encounters catastrophic plans during the training process, leading to sudden increases in WRL and GMRL. However, it can avoid these plans as training progresses, eventually converging and achieving relatively good results. Due to their simpler workload, their training processes converge in 1-3 hours and have low tail variance.

\subsubsection{Optimization Time}
\begin{figure}[tb]
  \centering
%   \subfigure[Optimization Time Comparison]
% {
    \begin{minipage}[b]{.49\linewidth}
        
        \centering
        \includegraphics[scale = 0.85]{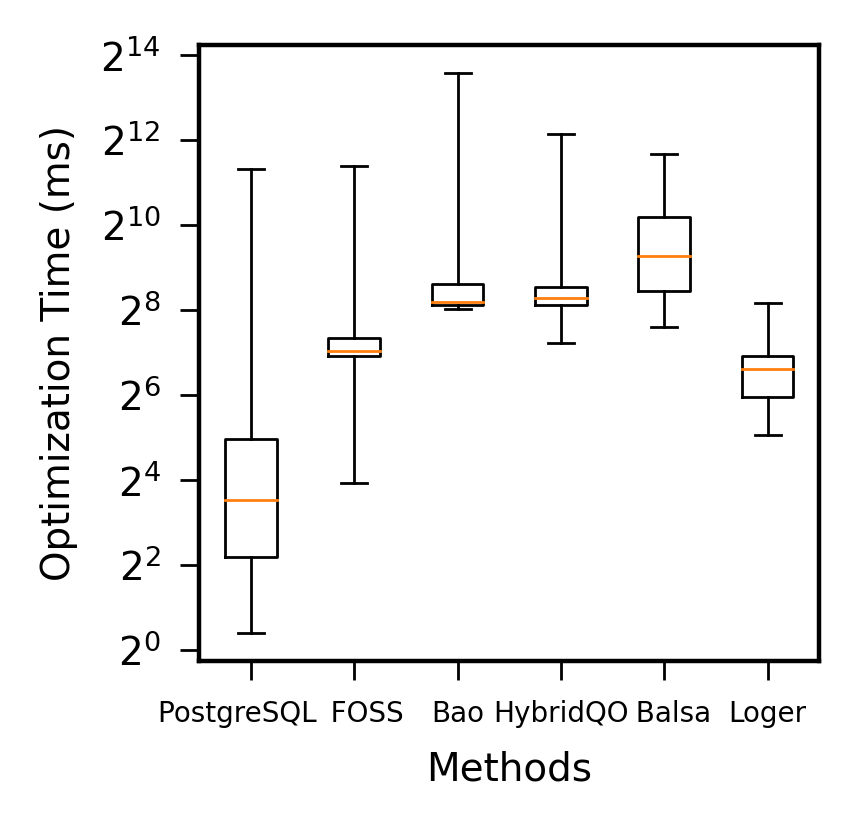}
        \caption{Optimization time of various optimizers.}
        \label{fig:optime}
    \end{minipage}
% }
%  \subfigure[]
% {
    \begin{minipage}[b]{.49\linewidth}
    
        \centering
        \includegraphics[scale = 0.85]{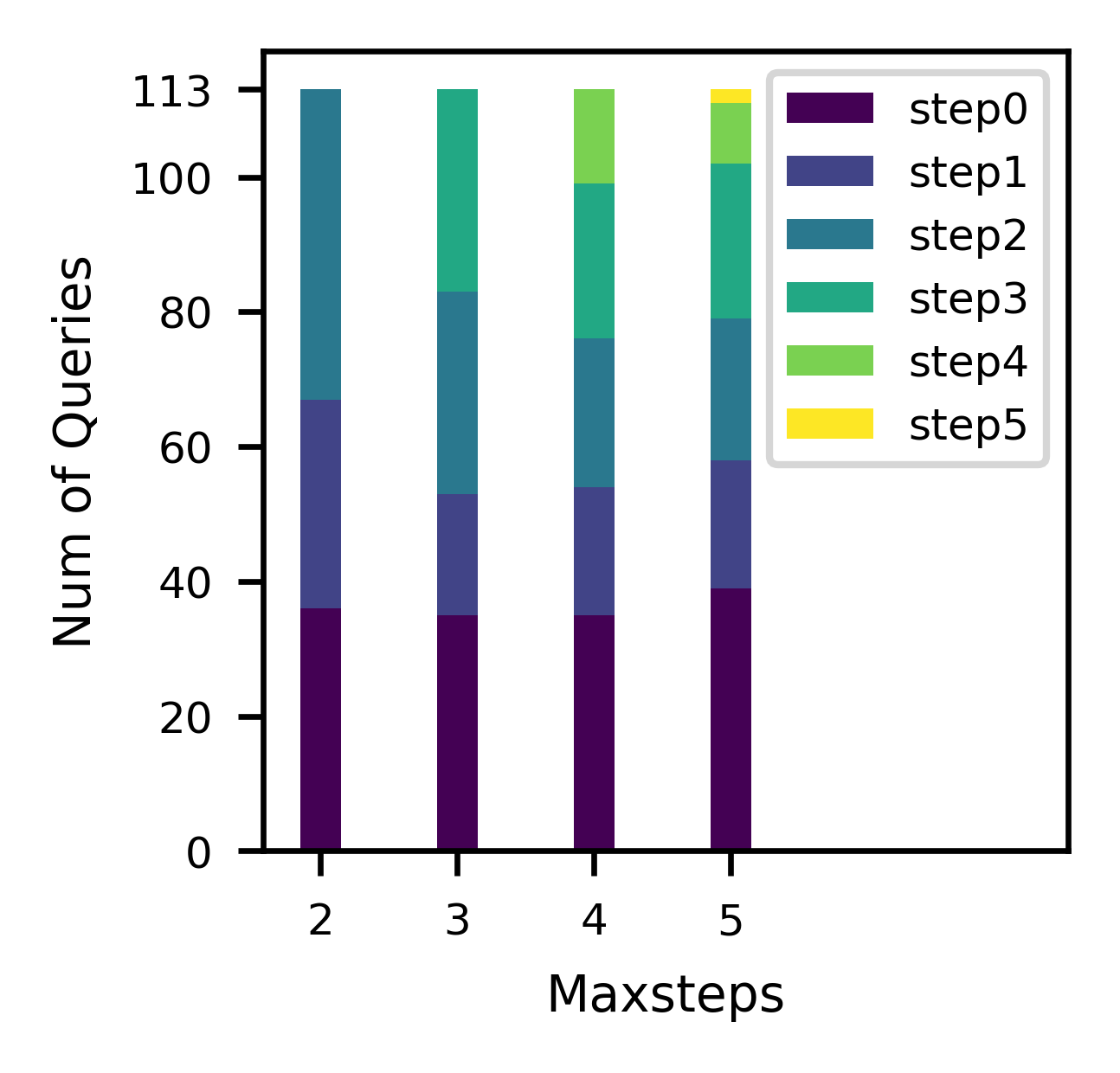}
        \caption{Steps distribution for known best plans under different maxsteps settings.}
        \label{fig:maxstepbuck}
    \end{minipage}
% }
\end{figure}

% \begin{table}[tb]% h asks to places the floating element [h]ere.
% \centering
% \caption{}
%   \label{tab:speedup}
%   \tabcolsep=0.05cm
%     \begin{tabular}{c|ccc|ccc}
%     % \cline{2-7}
%     \hline
%     % \  &Performance (WRL/GMRL)\\
%      \multirow{2}{*}{Methods}& \multicolumn{3}{c|}{Speedup}&\multicolumn{3}{c}{Training Time (Hours)} \\ 
%      \cline{2-7}
%       & JOB  & TPC-DS& STACK & JOB  & TPC-DS& STACK  \\
%      \hline
%     PostgreSQL& 2.78&  1.30&1.18& /& /&/\\
%     Bao & 1.22&  1.01&1.61& 3.35&  1.32&3.03\\
%     Balsa & 1.25&  2.42&NA& 12.18& 10.58&TLE\\
%     HybridQO& 1.42&  1.19&1.01& 3.51&  0.52&1.58\\
%     FOSS & 1.00&   1.00&1.00& 8.09&  1.55&2.31\\
%    \hline
% \end{tabular}
% \end{table}
We compare the optimization time of various optimizers on the entire JOB benchmark, which is the time from the input of SQL to the generation of the execution plan. As depicted in Fig. \ref{fig:optime}, we represent the comparative results using box plots. In contrast to PostgreSQL, learned-based optimizers often require longer optimization time due to operations like model inference and plan encoding. However, the overhead is negligible compared to the time saved in execution. Compared to Bao, Balsa, and HybridQO, FOSS has the smallest 25th percentile, 50th percentile, and 75th percentile of optimization time. The longer optimization time of FOSS compared to Loger is because the time taken by the traditional optimizer to generate the original plans accounts for a significant proportion of the optimization time in FOSS, while Loger does not require this. However, despite Loger having an advantage in optimization time, FOSS still surpasses it in total workload runtime.

% Excessive reliance on the expert optimizer results in long optimization time for Bao and HybridQO. Additionally, Balsa, characterized by a lengthy average decision sequence and a wide range of plan exploration strategies, also performs poorly in terms of optimization time. 
%Furthermore, there are many opportunities for engineering optimizations in the FOSS implementation that could further reduce the optimization time.
%However, the maximum optimization time of Balsa is even less than that of PostgreSQL. This is attributed to its minimal dependence on expert optimizer compared to the other three methods, allowing it to autonomously construct more comprehensive plans. 
\begin{figure}[tb]
  \centering
  \includegraphics[width=\linewidth]{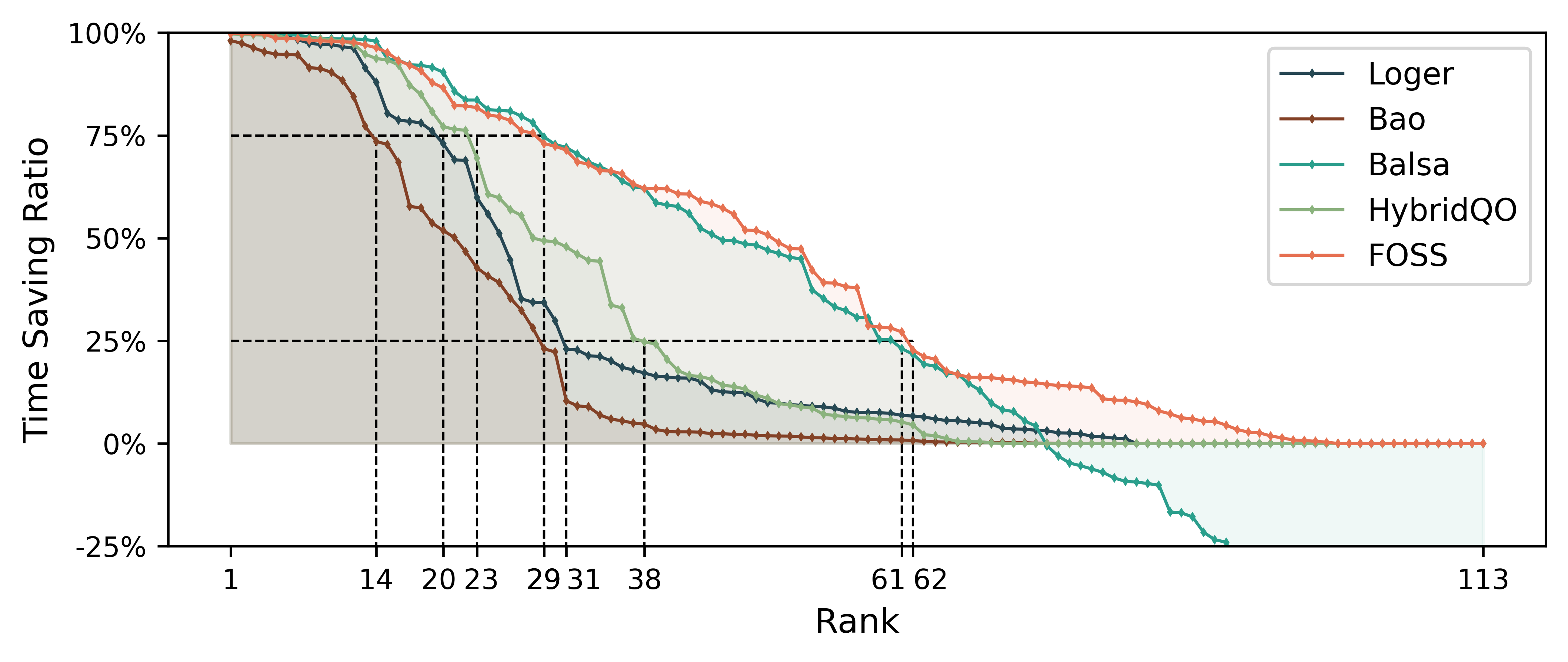}
  \caption{Ranking time savings ratios for known best plans.}
  \label{fig:KBPC}
\end{figure}
\subsubsection{Known Best Plan}
In this subsection, we analyze the maximum optimization potential for each learned query optimizer. We conduct three runs for each method on the entire JOB workload. From all the results, we acquire the known best plan (i.e., the one with the lowest execution latency) on each query for each learned optimizer. The comparison results are depicted in Fig. \ref{fig:KBPC}. It's sorted by the time savings ratios of the known best plans relative to the original plans.

Overall, FOSS comprehensively outperforms Bao, Loger, and HybridQO, surpassing the expert optimizer across a wider range of queries at every percentile. Due to the limited plan search space of Bao, it generates better-performance plans than the expert optimizer across only a minimal range of queries. Despite Balsa implementing more fine-grained optimization by considering additional table scan operators, FOSS can still exhibit similar performance to Balsa on the top 65 ranked queries. This suggests that FOSS, even when focusing solely on join order and join methods optimization, achieves significant optimization effects. It also can be observed that due to the lack of assurance from the original plan, Balsa generates plans with poorer performance than the expert optimizer on 40 queries. Specially, FOSS, Balsa, HybridQO, Loger, and Bao achieve at least a 25\% time savings on 62, 61, 38, 31 and 29 queries, respectively. They each save at least 75\% of the time on 29, 29, 23, 20 and 14 queries, respectively.
\label{Knwon Best Plan}
\subsection{Analysis of design choices}
\begin{figure}[tb]
 \centering
 	\includegraphics[width = \linewidth]{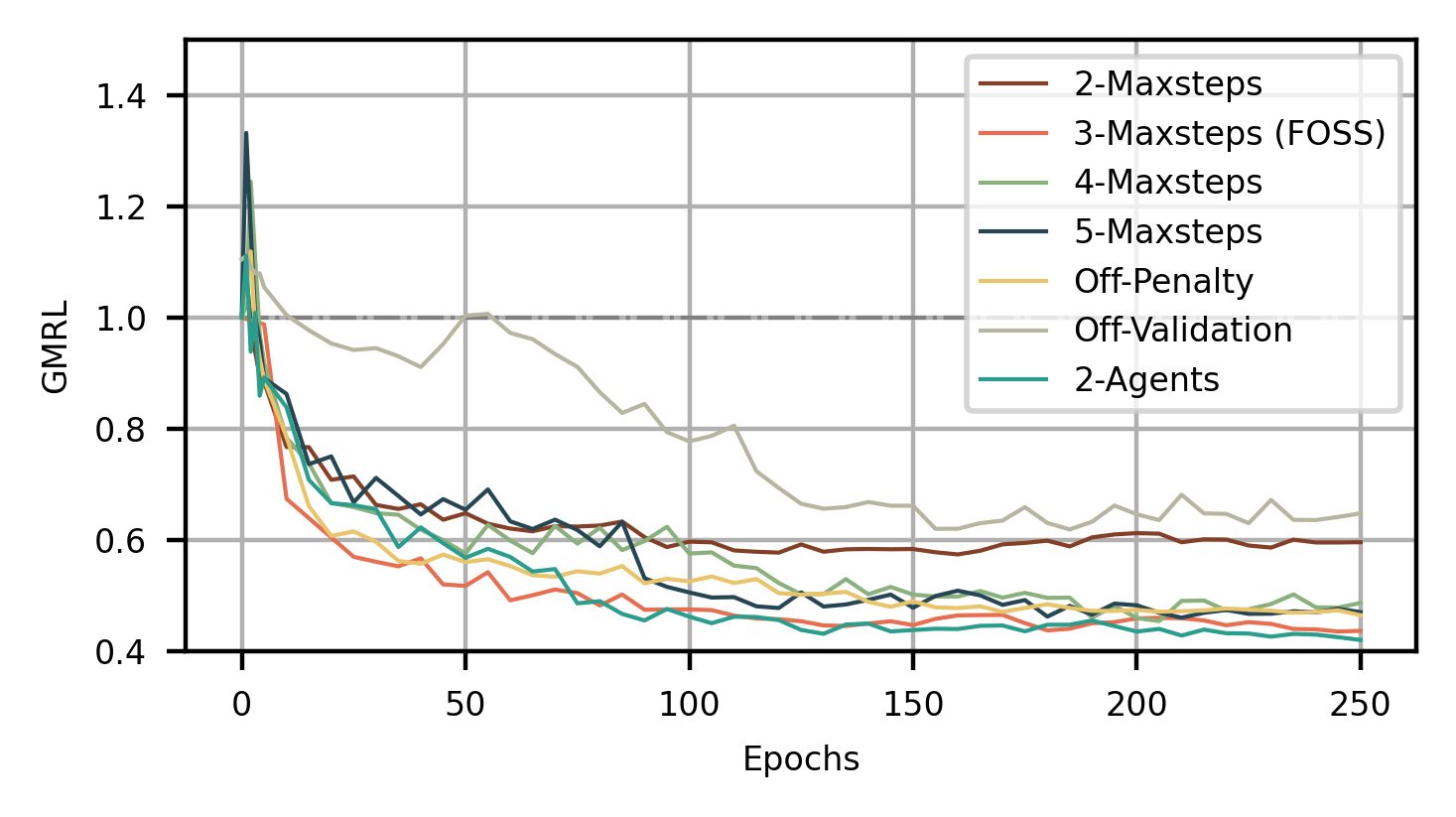}
\caption{GMRL variation curves under different configurations.}
\label{fig:maxsteprun}
\end{figure}

In this subsection, we analyze design choices in FOSS by individually replacing each component with alternatives. We test their performances on the entire JOB workload, using GMRL as the metric for evaluation, and consider training time and average plan optimization time for practical insights. Table \ref{tab:maxtable} presents our comparison results and Fig. \ref{fig:maxsteprun} illustrates the variation of GMRL during the training process.
\begin{table}[tb]% h asks to places the floating element [h]ere.
\centering
  \caption{Different Configurations Comparison Results.}
  \label{tab:maxtable}
  \tabcolsep=0.05cm
  \begin{tabular}{c|ccc}
     \hline
     \multirow{2}{*}{Experiments}& Training time& Optimization Time&  \multirow{2}{*}{GMRL}\\
      &  (hours)& (milliseconds)& \\
     \hline
        2-Maxsteps& 8.02& 182.77&0.596\\
        3-Maxsteps (FOSS)& 9.09& 209.01&0.436\\
        4-Maxsteps& 10.37& 235.14&0.487\\
        5-Maxsteps& 10.75& 259.45&0.470\\
        Off-Simulated& 48.01& 210.47&0.691\\
        Off-Penalty& 8.65& 208.29&0.465\\
        Off-Validation& 6.26& 209.82&0.653\\
        2-Agents& 12.45& 280.65&0.420\\
        % Retrain& 1.72& 172.15&0.44\\
     \hline
\end{tabular}
\end{table}
\subsubsection{Determination of Maxsteps}
Larger $maxsteps$ implies that FOSS can explore more diverse plans through the modifications of longer sequences to the original plan. However, it also makes the training of FOSS more challenging. We set $maxsteps$ to 2, 3, 4, and 5, respectively, and compare their performances. As described in Table \ref{tab:maxtable}, with an increase in $maxsteps$, the optimization time grows due to the consideration of more plans in one episode. Concurrently, training time gradually increases as the plan search space expands, and the number of plans to be executed grows. As illustrated in Fig. \ref{fig:maxsteprun}, when $maxsteps$ is set to 3, FOSS achieves the best performance with the minimum GMRL. 

For a more in-depth comparison, as described in Fig. \ref{fig:maxstepbuck}, we analyze the distribution of step status for known best plan, where \textit{step0} plans refers to the original plans, and \textit{stepN} plans refers to plans that take $N$ steps from the original plans. In the bar for $2-maxsteps$, there is a significant number of \textit{step2} plans. It suggests that $2-maxsteps$ may be insufficient. In the bar for $5-maxsteps$, the proportion of \textit{step4} and \textit{step5} plans is relatively small. As demonstrated by the other two results, FOSS always yields effective plans within 1-3 steps. 

The above results can be attributed to the fact that as $maxsteps$ increases, the longer action sequences complicate the agent's exploration and make it more challenging for the AAM to select a good plan from the candidate plans. Therefore, considering both training complexity and experimental effectiveness, we set the $maxsteps$ to 3.

% \textcolor{modify}{Despite spending more training time and optimization time than $3-maxsteps$, $5-maxsteps$ yields larger GMRL}.
\subsubsection{Effect of Simulated Environment}
To illustrate the effect of the simulated environment, we disable it in FOSS. It allows the agent to interact only with the real environment. Due to the extensive queries executed in the real environment, it significantly prolong the training time. To make training more feasible, we have to reduce the episode number for each agent update to 200. After 48 hours of training, despite benefiting from more reliable rewards feedback, its performance remains unsatisfactory. Its final GMRL is 0.691, which is significantly poorer than FOSS. The inadequacy can be ascribed to the insufficient exploration. It indicates the effectiveness of the simulated environment.
\subsubsection{Effect of Penalty}
Penalty measure in the reward function can guide FOSS to focus on generating valuable candidate plans. We set penalty coefficient to 0 in the default configuration. Table \ref{tab:maxtable} illustrates that FOSS without penalty, although having a shorter training time compared to FOSS with penalty, performs worse. This can be attributed to FOSS with penalty allowing the planner to generate a more diverse set of candidate plans within a limited number of steps. It also enriches the training data for updating the AAM. This indicates that penalty measure has a positive impact on FOSS.
% FOSS with penalty allowing the planner to generate candidate plans in as few steps as possible, resulting in a greater diversity of candidate plans generated within the limited steps. 
\subsubsection{Effect of Promising Plans Validation}
During planner training, FOSS collects the promising plans based on the feedback from the AAM and execute these plans when compute resources are idle. Then FOSS pushes their execution results into the execution buffer for updating the AAM. In this experiment, we remove this operation, meaning FOSS can only obtain training data by randomly sampling queries and collecting candidate plans. As described in Table \ref{tab:maxtable}, it reduces training time but diminishes the performance. As shown in Fig. \ref{fig:maxsteprun}, its GMRL decreases slowly with training progress. The absence of validation for promising plans not only reduces diversity in the AAM training data, but also leads to the accumulation of estimation errors during planner training, as errors in the AAM are not corrected in a timely manner.
\subsubsection{Training with Muti-Agents}
To enhance the robustness of FOSS, we also incorporate a switch for multi-agents. FOSS can initialize $m$ agents with different strategies (e.g., different discount factors and learning rates). In training mode, they interact with the real environment, obtaining execution result. These data are then integrated into the execution buffer and employed to train the AAM. Subsequently, they individually interact with the simulated environment to generate simulated experiences and concurrently update themselves. In inference mode, for a given query, $m$ agents independently generate candidate plans, utilize AAM to select the best plan, and the ultimately execution plan is chosen from the $m$ best plans using AAM. 

We deploy FOSS with 2 agents. It demonstrates better performance than FOSS with 1 agent (Table \ref{tab:maxtable}) and requires fewer epochs for convergence (Fig. \ref{fig:maxsteprun}). This can be attributed to the strategy of multi-agents, which not only enriches the diversity of training data for AAM but also generates a more diverse set of high-quality candidate plans. However, due to our serialization configuration, it incurs more training time than FOSS with 1 agent. If computational resources are ample, we recommend deploying FOSS with multi-agents in parallel.
% \subsubsection{Efficiency of Retraining Agent}
% \label{chapter_retrain}
% For FOSS, in the presence of a well-performing advantage model, the cost of retraining an new agent with a completely new strategy (such as modifying the agent's network or parameters) is minimal. As shown in Table \ref{tab:maxtable}, by engaging in multiple epochs of interaction with the simulated environment for just about two hours, the new agent can acquire optimization capabilities that match or even surpass that of the original agent. The new agent can also be utilized to assist in optimizing the original plans, thereby enhancing the robustness of FOSS.
\section{Related Work}
\noindent\textbf{Learned Query Optimizer.} In recent years, various learned query optimizers have been proposed. Apart from the two types of methods introduced in \ref{Intro}, there are also other works aiming to replace the traditional cost model and incorporate appropriate plan search space reduction methods. For instance, Robopt \cite{Robopt} encodes plans to predict plan performance for cross-platform systems, basing the entire plan enumeration on a set of algebraic operations that operate on vectors. Leon \cite{Leon} retains expert knowledge as much as possible. It follows the plan generation process of the expert optimizer and only replaces the traditional cost model with a learned calibration model initialized by the traditional cost model. Additionally, Leon employs another learned model to help reduce unnecessary plan space. These two learned models effectively balance immediate and long-term benefits for plan generation. Similarly, QPSeeker \cite{QPSpeeker} also aims to replace the traditional cost model with a powerful learned model that learns from both data distribution and queries with the assistance of a language model. Additionally, it utilizes Monte Carlo Tree Search to provide an execution plan for the query.

There are also some works \cite{AutoSteer,FASTgres} build upon the foundation laid by Bao\cite{Bao}, employing hint sets similar to Bao. They focus on making the selection of candidate hint set more intelligent rather than relying solely on expert knowledge. Their work contributes positively to the advancement of plan-steerer.\\
\textbf{Reinforcement Learning.} Reinforcement learning can be divided into model-free RL and model-based RL. The former have experienced rapid development in the past few years, with many classic algorithms \cite{DQN,DDQN,PPO,A3C,SAC} being proposed and demonstrating excellent performance in various fields \cite{DRL1,DRL2,DRL3,DRL4}. However, model-free RL with high sample complexity are hard to be directly applied in real-world tasks, where trial-and-errors can be highly costly \cite{surveymbrl}. As a result, some scholars are dedicated to building simulated environment to interact more efficiently with agents. FOSS follows the Dyna-style \cite{dyna} method, using the learned environment model to generate more simulated experiences for the agent. In addition to the mentioned approach, in many model-based RL scenarios, the simulated environment models are differentiable. This enables policy learning through differential planning and value gradient methods \cite{surveymbrl}.
% \textbf{Reinforcement Learning From Human Feedback.} Reinforcement learning from human feedback (RLHF) is a technique employed for training AI systems to align with human goals. RLHF has emerged as the primary method for fine-tuning SOTA large language models \cite{RLHFsurvey}. The standard methodology for RLHF used today was introduced by \cite{RLHF} in 2017. Its framework is similar to FOSS, requiring feedback from external comparisons of two samples to train a reward model. This reward model is then used to train traditional RL, which is policy optimization. The key difference is that RLHF receives feedback from humans, making sample labeling more challenging due to human subjectivity, whereas FOSS obtains feedback from DBMS's real execution.
\section{Conclusion}
We introduce FOSS, a novel DRL-based query optimization system. Starting from the original plans generated by traditional optimizer, the planner of FOSS experiments with the optimization of these plans through step-by-step modifications, which results in candidate plans. We design an asymmetric advantage model to serve as a selector, choosing the optimal one from the candidate plans. Furthermore, to enhance training efficiency, we employ the asymmetric advantage model as a reward indicator and integrate it with a traditional optimizer to build a simulated environment. With the simulated environment, FOSS can enhance its optimization ability through self-bootstrapping.

To our knowledge, FOSS is the pioneering learned-based query optimizer that achieves better-performance plans through fine-grained modifications to the original plans and efficiently self-bootstraps by interacting with the simulated environment model. We believe that FOSS opens up a new path for query optimization and deserves further exploration.
\section*{Acknowledgment}
National Key Research \& Develop Plan(2023YFB4503600); National Natural Science Foundation of China (U23A20299, 62072460, 62172424, 62276270, 62322214).

\bibliographystyle{IEEEtran}
\bibliography{IEEEabrv,sample}

%\end{thebibliography}

% \vspace{12pt}
% \color{red}
% IEEE conference templates contain guidance text for composing and formatting conference papers. Please ensure that all template text is removed from your conference paper prior to submission to the conference. Failure to remove the template text from your paper may result in your paper not being published.

\end{document}